\documentclass[twocolumn]{aastex63}

\usepackage{graphicx} 
\usepackage{float} 
\usepackage{natbib}
\usepackage{amsmath}
\usepackage[T1]{fontenc}
\usepackage[utf8]{inputenc}

\def\red#1 {\textcolor{red}{#1}\ } 

\shorttitle{HEM Perturber Observables}
\shortauthors{Jackson et al.}

\begin{document}
\title{Observable Predictions from Perturber-coupled High-eccentricity Tidal Migration of Warm Jupiters}

\author[0000-0002-0323-4828]{Jonathan M. Jackson}
\affil{Department of Astronomy \& Astrophysics, Center for Exoplanets and Habitable Worlds, The Pennsylvania State University, University Park, PA 16802}
\author[0000-0001-9677-1296]{Rebekah I. Dawson}
\affil{Department of Astronomy \& Astrophysics, Center for Exoplanets and Habitable Worlds, The Pennsylvania State University, University Park, PA 16802}
\author[0000-0002-0711-4516]{Andrew Shannon}
\affil{LESIA, Observatoire de Paris, Universit\'{e} PSL, CNRS, Sorbonne Universit\'{e}, Universit\'{e} de Paris, 5 place Jules Janssen, 92195 Meudon, France}
\affil{Department of Astronomy \& Astrophysics, Center for Exoplanets and Habitable Worlds, The Pennsylvania State University, University Park, PA 16802}
\author[0000-0003-0412-9314]{Cristobal Petrovich}
\affil{Steward Observatory, University of Arizona, 933 N. Cherry Ave., Tucson, AZ 85721, USA}
\affil{Pontificia Universidad Cat\'olica de Chile, Facultad de F\'{i}sica, Instituto de Astrof\'\i sica, Av. Vicu\~{n}a Mackenna 4860, 7820436 Macul, Santiago, Chile}
\affil{Millennium Institute for Astrophysics, Chile}
\correspondingauthor{Jonathan M. Jackson}
\email{jqj5401@psu.edu}

\begin{abstract}
The origin of warm Jupiters (gas giant planets with periods between 10 and 200 days) is an open question in exoplanet formation and evolution. We investigate a particular migration theory in which a warm Jupiter is coupled to a perturbing companion planet that excites secular eccentricity oscillations in the warm Jupiter, leading to periodic close stellar passages that can tidally shrink and circularize its orbit. If such companions exist in warm Jupiter systems, they are likely to be massive and close-in, making them potentially detectable. We generate a set of warm Jupiter-perturber populations capable of engaging in high-eccentricity tidal migration and calculate the detectability of the perturbers through a variety of observational metrics. We show that a small percentage of these perturbers should be detectable in the \textit{Kepler} light curves, but most should be detectable with precise radial velocity measurements over a 3-month baseline and \textit{Gaia} astrometry. We find these results to be robust to the assumptions made for the perturber parameter distributions. If a high-precision radial velocity search for companions to warm Jupiters does not find evidence of a significant number of massive companions over a 3-month baseline, it will suggest that perturber-coupled high-eccentricity migration is not the predominant delivery method for warm Jupiters.
\end{abstract}


\section{Introduction}\label{intro}

Both hot (periods < 10 days) and warm (periods between 10 and 200 days) Jupiters are commonly thought to have formed beyond the ice-line and migrated inward to their current semi-major axes (\citealt{Bod00,Raf06}; see \citealt{Daw18} for a review of hot Jupiter origins theories). Disk migration \citep{Gol80,War97,Bar14} could deliver hot Jupiters, but it is difficult to reconcile with the wide eccentricity distribution of warm Jupiters because planet-disk interactions tend to damp eccentricities \citep{Bit13,Dun13}. Planet-disk interactions can sometimes excite eccentricities, but they typically saturate at a random velocity equal to the sound speed ($e\lesssim 0.03$ for a 100 day orbit; \citealt{Duf15}). The eccentricity distribution is difficult to explain with post-disk planet-planet scattering because eccentricity growth via scattering is limited by $v_{\rm{escape}}/v_{\rm{keplerian}}$, which corresponds to an $e_{\rm{max}}\sim0.5$ for a 0.5 $m_{\rm{Jup}}$ 2 $r_{\rm{Jup}}$ planet on a 50 day orbit \citep{Gol04,Ida13,Pet14}. 

Warm Jupiters, particularly those with intermediate to large eccentricities, may have arrived at their short orbital periods (10--200 days) through high-eccentricity tidal migration (e.g., \citealt{Wu03}). In this scenario, the planets form at $\sim$a few AU and their orbits are excited to large eccentricities. Possible mechanisms of eccentricity excitation include planet-planet scattering \citep{Ras96}, secular chaos \citep{Wu11}, and stellar flybys (e.g., \citealt{Kai13}). These mechanisms are successful in producing a wide distribution of giant planet eccentricities from multi-planet systems with initially circular orbits (e.g., \citealt{Jur08,Cha08}). Once the planets are excited to sufficiently large eccentricities, tidal dissipation in the planet during close pericenter passages shrinks and circularizes their orbits. 

Tidal dissipation is strongly dependent on the distance from the star. Although a couple warm Jupiters such as HD 80606 b \citep{Nae01} and HD 17156 b \citep{Fis07} have pericenter distances potentially small enough for tidal migration, most eccentric warm Jupiters do not. One possibility is that warm Jupiters are undergoing Kozai-Lidov oscillations or other secular eccentricity oscillations induced by an exterior perturber. In this scenario, the planets spend a small percentage of their time at eccentricities large enough for tidal migration, but would typically be observed at more moderate eccentricities (e.g., \citealt{Tak05,Don14}). Eccentricity oscillations have been studied in some warm Jupiter systems with companions (e.g., \citealt{Kan14}), but they are dependent on the inclination which is often unknown. The presence of subjovian companions in many of these systems suggests the high-eccentricity migration model cannot account for all warm Jupiters \citep{Hua16}; However, it may still explain those with intermediate to large eccentricities. \citet{Pet16} determined that perturber-coupled high-eccentricity tidal migration can account for $\sim20\%$ of warm Jupiters and most warm Jupiters with $e\geq0.4$.

\citet{Don14} showed that warm Jupiters need massive, nearby companions to engage in perturber-coupled high-eccentricity tidal migration. They derived equations for the perturbing strength needed to overcome general relativistic precession and reach large eccentricities. In this paper, we construct a synthetic population of perturbers that meet these requirements and calculate their observational signatures (including transit timing variations, transit duration variations, radial velocities, and astrometry). In Section \ref{pop} we construct our synthetic warm Jupiter-perturber systems. In Section \ref{results} we calculate the observational signatures of the perturbers in those systems. In Section \ref{pop_comp} we assess the robustness of our choices of parameter distributions. In Section \ref{petro} we perform the same calculations using the perturber population from \citet{Pet16}. In Section \ref{companions} we summarize the population of observed warm Jupiters with massive exterior companions and compare these systems to our results. Lastly, we summarize our conclusions and interpret our results in Section \ref{conc}.

\section{Population Construction}\label{pop}

We construct a population of warm Jupiter-perturber systems capable of engaging in high-eccentricity tidal migration. We will calculate the observational signatures of this population of perturbers and compare them to current detection limits (Section \ref{results}). In the following two subsections, we describe how we build our population of warm Jupiters (Sections \ref{kepler_warm}, \ref{tess_warm}, and \ref{rv_warm}) and their corresponding perturbers (Section \ref{perturbers}) from initial parameter distributions that are informed by observed systems.

\subsection{\textit{Kepler}-like Warm Jupiter Population}\label{kepler_warm}

Before we can construct the population of perturbers and assess their observability, we must first define a population of warm Jupiters. The properties of the warm Jupiters will affect the stability of the systems and the capability of the warm Jupiter to undergo large secular eccentricity oscillations despite general relativistic precession, both of which will dictate the requirements for the perturbers (see Section \ref{perturbers}). We assume for the purpose of constructing this population that all warm Jupiters -- even those observed with low eccentricities -- underwent perturber-coupled high-eccentricity migration in order to reach their current orbits.

\begin{figure}
\center{\includegraphics{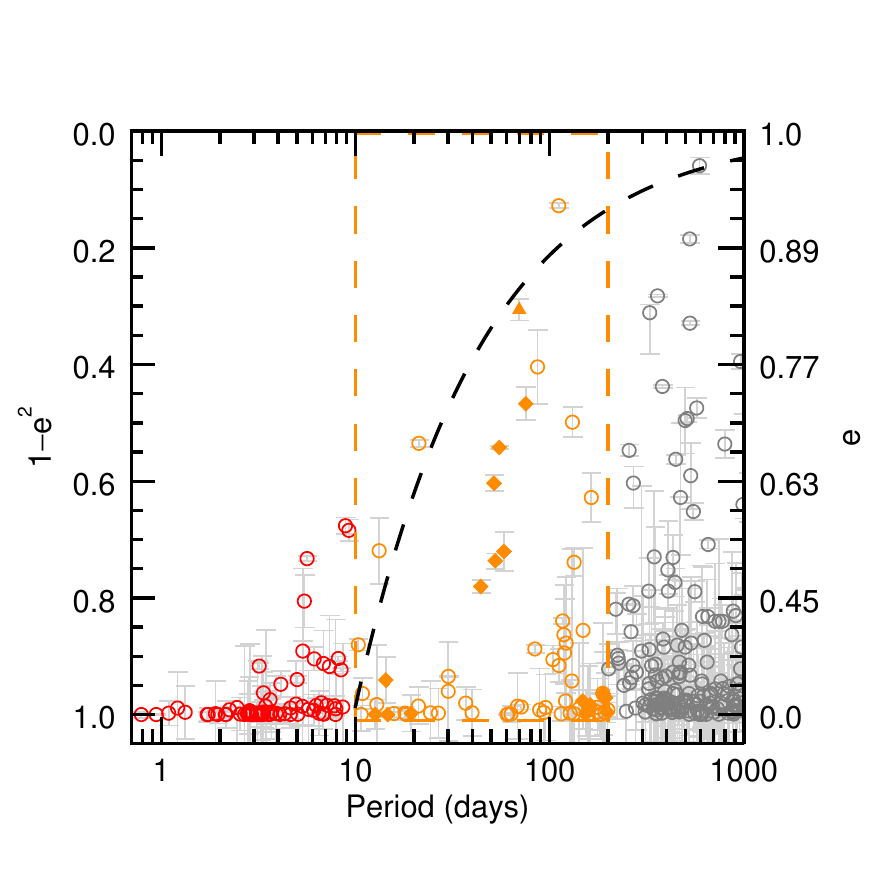}}
\caption{Here we highlight the warm Jupiter population of exoplanets (orange) in the context of hot (red) and cold (gray) Jupiters. Each point on this plot represents an exoplanet with $m>0.5m_{\textrm{Jup}}$ and measured eccentricity. Among the warm Jupiters, open circles represent planets with no known companions, closed diamonds represent planets with companions detected via radial velocities, and closed triangles represent planets with companions detected via transit timing variations. The black dashed line represents the tidal circularization track of constant angular momentum that a migrating warm Jupiter would follow if it were to end its migration as a circular hot Jupiter on a 10-day orbit. Data taken from the NASA Exoplanet Archive (as of 2020 February 26).}
\label{fig:ecc}
\end{figure}

In Figure \ref{fig:ecc}, we show the observed period and eccentricity distribution of Jupiter-sized planets. Hot Jupiters are plotted in red, warm Jupiters in orange, and cold Jupiters in gray. Warm Jupiters with companions are plotted as diamonds (companion discovered with radial velocities) or triangles (companions discovered with transit timing variations). These planets will be discussed further in Section \ref{companions}. The black dashed line is a track of constant angular momentum and will be discussed further in Section \ref{perturbers}. Note that the broad distribution of warm Jupiter eccentricities seen here is one of the motivators for researching high-eccentricity migration mechanisms.

We base our warm Jupiter population on characteristics of the observed distribution. We begin by creating a \textit{Kepler}-like sample of transiting warm Jupiters which will be appropriate for assessing the observability of transit timing and duration variations. These planets have periods between 10 and 200 days and masses between 0.1 and 10 $m_{\rm{Jup}}$. To build this population, we draw the planet's period, mass, and eccentricity from distributions that model the intrinsic properties of warm Jupiters and then discard any planets with impact parameter $b<1$, which would not transit.

We draw our warm Jupiter periods from a power law with fitting coefficient $P=0.63$ \citep{Fer19}. Since the primary Kepler mission lasted $\sim 4$ years and 3 transits were required to detect a planet, the period distribution is complete out to the 200 day maximum in our sample.

We draw our warm Jupiter eccentricities from a beta distribution (e.g., \citealt{Kip13}). To find the appropriate fitting parameters for warm Jupiters, we fit a beta distribution to the sample of 102 confirmed planets on the NASA Exoplanet Archive with periods between 10 and 200 days and masses $>0.1 m_{\rm{Jup}}$ discovered via the radial velocity technique (as of 2020 February 26). The functional form of this distribution is

\begin{equation}
\label{eq:beta}
    P(e;\alpha,\beta)=\frac{\Gamma(\alpha+\beta)}{\Gamma(\alpha)\Gamma(\beta)}e^{\alpha-1}(1-e)^{\beta-1},
\end{equation}

where $e$ is the eccentricity, $\Gamma()$ is the gamma function, and $\alpha$ and $\beta$ are fitting parameters. Our best fit parameters are $\alpha=0.61$ and $\beta=2.16$. We discard any planets whose semi-major axis and eccentricity would result in tidal disruption: $a(1-e)<a_{\rm{Roche}}$, where $a_{\rm{Roche}}$ includes a scaling coefficient of $f_{\rm{p}}=2.7$ (e.g., \citealt{Gui11}). We note that some planets with eccentricities just below this limit may not survive as warm Jupiters for long due to rapid tidal circularization, but this is strongly dependent on the particular tidal parameters of the system.

We draw our warm Jupiter inclinations from an isotropic distribution (i.e., uniform in cos$i$). We note, however, that after the cut for transiting planets, the inclination distribution will be strongly peaked near $\pi/2$.

Lastly, we assume a power law mass distribution ($ dN/d\rm{ln}(m)\propto m^{\alpha}$) with coefficient $\alpha = -0.31$ found by \citet{Cum08}. These authors considered selection effects in the RV sample to estimate the mass distribution of nearby giant planets. We limit our masses to be between 0.1 and 10 $m_{\rm{Jup}}$.

To finish constructing our \textit{Kepler}-like warm Jupiter population, in addition to the periods ($P$), eccentricities ($e$), inclinations ($i$), and planet masses ($m_{\rm{wj}}$) described above, we draw arguments of pericenter ($\omega$), and mean anomalies ($M$) from uniform distributions before applying our impact parameter cut of $b<1$. It is not necessary to cut in orbital period or eccentricity to account for detectability because the \textit{Kepler} is complete for all eccentricities and periods < 200 days for the mass range we are considering. For each system, we assume a solar mass and radius star. Through this method, we build a population of $10,000$ warm Jupiter-star systems.

\subsection{\textit{TESS}-like Warm Jupiter population} \label{tess_warm}

In addition to transit timing variations (TTVs) and transit duration variations (TDVs), we want to assess the detectability of our perturbers via radial velocities (RVs) and astrometry. Unfortunately, most \textit{Kepler} systems are too distant for RV or astrometric followup, which can reliably detect planets around bright stars (V $\lesssim 12$; see \citealt{Per14}). However, \textit{TESS} systems and warm Jupiters discovered by the RV method (see Section \ref{rv_warm}) are typically much brighter and closer, making them more amenable for followup. Thus, here we define a population of \textit{TESS}-like warm Jupiter systems.

Our \textit{TESS}-like sample is very similar to the \textit{Kepler} sample defined in Section \ref{kepler_warm}, with the exception of its period distribution. Since \textit{TESS} is an all-sky mission, individual stars will be observed for a much shorter period of time than the equivalent from the \textit{Kepler} mission, making the expected sample incomplete to our 200 day maximum period for warm Jupiters. To account for this incompleteness, we take the periods of planets with radius $r>8R_{\rm{Earth}}$ from a simulated \textit{TESS} yield \citep{Bar18}, drawing a period value at random from this list for each iteration in our warm Jupiter population construction process. This simulated \textit{TESS} yield approximates the distribution of expected planetary parameters in the \textit{TESS} sample based on current planet occurrence rate estimates and the \textit{TESS} instrumental capabilities. \citet{Bar18} predicts the discovery of roughly 300 warm Jupiters by \textit{TESS}.

We follow the method described in Section \ref{kepler_warm} to draw eccentricities ($e$), inclinations ($i$), masses ($m_{\rm{wj}}$), arguments of pericenter ($\omega$), and mean anomalies ($M$), applying the same cuts for transiting planets and stable warm Jupiter orbits. As with the \textit{Kepler} planets, we build a population of 10,000 \textit{TESS}-like warm Jupiter systems. 

\subsection{RV-discovered warm Jupiter population} \label{rv_warm}

Lastly, we would like to assess the detectability of perturbing companions to warm Jupiters discovered by the radial velocity (RV) method. These companions could be detectable by RVs or astrometry because the host stars are typically bright and nearby.

Our population of RV-detected warm Jupiters is more straightforward to develop than the transiting populations described in Sections \ref{kepler_warm} and \ref{tess_warm} because the masses and eccentricities are well-characterized. Thus, we can directly use the observed planets, pulling their minimum masses ($m_{\rm{wj}}\sin{i_{\rm{wj}}}$), periods ($P$), eccentricities ($e$), arguments of pericenter ($\omega$), and stellar mass ($m_*$) from the sample of 66 RV-detected warm Jupiters with well-characterized orbits on the NASA Exoplanet Archive (as of 2020 February 26). We draw the mean anomalies ($M$) of the planets from a uniform distribution and the inclinations ($i$) from an isotropic distribution, cutting out orientations in which the derived $m_{\rm{wj}}$ is above 10 $m_{\rm{Jup}}$. The use of an isotropic inclination distribution introduces a slight bias because we are now dealing with observed planets instead of an intrinsic population; however, this effect is small for the mass range we are considering \citep{Ho11,Lop12}. We draw from our sample of warm Jupiters 10,000 times to create a population similar to our two transit-detected warm Jupiter populations.

\subsection{Perturber Population}\label{perturbers}
Now we construct the population of perturbing companion planets that will each accompany a warm Jupiter. We set two criteria for accepting a given perturber: (1) the warm Jupiter-perturber system must be long-term stable and (2) the perturber must be capable of periodically exciting the eccentricity of the warm Jupiter high enough for it to undergo tidal migration through an ongoing cycle of secular angular momentum exchange.

To implement the first criterion, we impose an analytical stability requirement from \citet{Pet15b} which, when satisfied, should result in the system evolving secularly with no close encounters:

\begin{equation}
\label{eq:petro}
    \frac{a_{\rm{per}}(1-e_{\rm{per}})}{a_{\rm{wj}}(1+e_{\rm{wj}})}>2.4[\max(\mu_{\rm{wj}},\mu_{\rm{per}})]^{1/3}\left(\frac{a_{\rm{per}}}{a_{\rm{wj}}}\right)^{1/2}+1.5,
\end{equation}

where $a_{\rm{per(wj)}}$ is the semi-major axis of the perturber(warm Jupiter) planet, $e_{\rm{per(wj)}}$ is the eccentricity of the perturber(warm Jupiter) planet, and $\mu_{\rm{per(wj)}}=m_{\rm{per(wj)}}/m_*$ is the perturber(warm Jupiter) planet-to-star mass ratio.

To implement the second criterion, we first set an empirical limit on the warm Jupiter's minimum angular momentum for tidal migration:

\begin{equation}
\label{eq:limit}
    a(1-e^2)<a_{\rm{f,crit}}=0.1\; \rm{AU},
\end{equation}

where $a(1-e^2)$ is the final semi-major axis of a tidally circularized planet when $e=0$ and $a_{\rm{f,crit}}=0.1$ AU is the final semi-major axis required to circularize within the age of the universe. The value of $a_{\rm{f,crit}}$ comes from the observed population of hot Jupiters which tend to exhibit small eccentricities for $a<0.1$ AU \citep{Soc12,Don14}. In reality, the limit is likely much stricter since there are still some moderate eccentricities among hot Jupiters with $a>0.05$ and the tidal circularization timescale scales vary strongly with separation.  However, we adopt the conservative value of $a_{\rm{f,crit}}=0.1$ AU to ensure the criterion presented below is a necessary requirement for perturber-coupled high-eccentricity tidal migration. Figure \ref{fig:ecc} shows the theoretical limit from equation \ref{eq:limit} plotted as a black dashed line over the observed Jupiter-mass planet $a$ vs. $e$ distribution. A planet must spend time above the line on this plot in order to tidally migrate to become a hot Jupiter.

We impose an analytical requirement from \citet{Don14} which, when satisfied, implies that the perturber is capable of inducing eccentricity oscillations in the warm Jupiter that can overcome precession due to general relativity and satisfy equation \ref{eq:limit} for some portion of its secular cycle. This is a minimum requirement for high-eccentricity tidal migration, but does not guarantee it will occur.

\begin{eqnarray}
\label{eq:dong}
    \frac{b_{\rm{per}}}{a_{\rm{wj}}}<\left(\frac{8Gm_*}{c^2a_{\rm{wj}}}\right)^{-1/3}\left(\frac{m_*}{m_{\rm{per}}}\right)^{-1/3} \nonumber \\
    \times\left(\sqrt{\frac{a_{\rm{wj}}}{a_{\rm{f,crit}}}}-\frac{1}{\sqrt{1-e_{\rm{wj}}^2}}\right)^{-1/3}\left[1-\frac{a_{\rm{f,crit}}}{a_{\rm{wj}}}-e_{\rm{wj}}^2\right]^{1/3} \nonumber \\
    \times\left(2-5\sin^2i_{\rm{mut}}\sin^2\omega_{\rm{wj}}\right)^{1/3}, \nonumber \\
\end{eqnarray}

where $b_{\rm{per}}=a_{\rm{per}}(1-e_{\rm{per}}^2)^{1/2}$ is the semi-minor axis of the perturber, $m_*$ is the stellar mass, $m_{\rm{per}}$ is the mass of the perturber, $e_{\rm{wj}}$ is the eccentricity of the warm Jupiter, $a_{\rm{wj}}$ is the semi-major axis of the warm Jupiter, $\omega_{\rm{wj}}$ is the argument of pericenter of the warm Jupiter in the invariable plane, $i_{\rm{mut}}$ is the mutual inclination between the two planets, and $a_{\rm{f,crit}}=0.1$.

The population of planets with orbital periods beyond 200 days is less constrained by the observations than that of the warm Jupiters, so we have to make some assumptions about the distribution of orbital and physical parameters. The robustness of our choices for these distributions is discussed in Section \ref{pop_comp}.

Our population of perturber eccentricities ($e$) is drawn from a Beta distribution (equation \ref{eq:beta}) with best fit parameters $\alpha=0.74$ and $\beta=1.61$. determined by fitting the beta distribution to the sample of 428 long-period, Jupiter sized ($m_{\rm{Jup}}>0.1$) confirmed planets discovered via radial velocity from the NASA Exoplanet Archive (as of 2020 February 26).

The perturber sky-plane inclinations ($i$) are drawn from an isotropic distribution. The orbital angles $\omega$, $\Omega$, and $M$ are drawn from uniform distributions in the interval $[0,2\pi]$. For the perturber population, we draw the masses ($m$) from the power law distributions calculated by \citet{Cum08} with a range of 0.1 to 20 $m_{\rm{Jup}}$. We choose the range of masses based on previous studies that show consistency with the power law fit in that range (e.g., \citealt{Fer19}). Above 20 $m_{\rm{Jup}}$, companions are more likely to be a part of the low-mass end of the brown dwarf population, which follows a different mass function \citep{Bow16}. We assess our choice for the upper mass limit in Section \ref{mass}.

Our perturber periods follow a symmetric broken power law,
\begin{equation}
\label{eq:broken}
    \frac{dN}{d\textrm{log}P}\propto \begin{cases} 
      \left(\frac{P}{P_{\textrm{break}}}\right)^{P_1} & P\leq P_{\textrm{break}} \\
      \left(\frac{P}{P_{\textrm{break}}}\right)^{-P_1} & P> P_{\textrm{break}}
   \end{cases}
\end{equation}
where $P_{\textrm{break}}=859 \textrm{days}$ is the turnover point in the period distribution and $P_1=0.63$ is the power law coefficient. \citet{Fer19} showed that a broken power law (with these fitting coefficients) is a better fit to the observed period distribution than a single power law or a log normal distribution. We draw our periods from the range 200 to 100,000 days. The lower limit is set to ensure these are exterior companions to warm Jupiters and the upper limit is set by the fact that planet-mass objects with periods beyond 100,000 days always fail equation \ref{eq:dong} for the mass range we test.

For a given warm Jupiter-star system, we draw perturber properties from the distributions listed above and check against our two acceptance constraints (equations \ref{eq:petro} and \ref{eq:dong}). If the set of properties fails either of the criteria, we redraw the parameters and recheck against the criteria. We repeat this process until every warm Jupiter has an accompanying perturber.

\section{Observational signatures of perturbers to warm Jupiters}\label{results}

In this section, we compute the observational signatures for our sample of synthesized systems (Section \ref{pop}). For each warm Jupiter discovery method (i.e., \textit{Kepler} transits, \textit{TESS} transits, radial velocities), we have 10,000 systems on which to run our calculations of perturber observables. In Sections \ref{ttv}, \ref{tdv}, \ref{rv}, and \ref{astro}, we calculate the transit timing variation (TTV), transit duration variation (TDV), radial velocity (rv), and astrometric signals, respectively. Finally, in Section \ref{detect}, we assess the overall detectability of our sample of perturbers.

\subsection{Transit Timing Variations}\label{ttv}

First, we calculate the transit timing variation (TTV) signal of the perturbers to the \textit{Kepler}-like warm Jupiters. We do not run this calculation on the \textit{TESS}-like systems because the time baselines of most \textit{TESS} planets will be too short to detect TTVs. The type of TTVs we expect are typical of those seen in hierarchical triple systems (e.g., \citealt{Bor03,Bor04}).

We model the TTVs over the \textit{Kepler} mission lifetime of 4 years. We calculate the TTVs numerically by simulating the 3-body motion of the two planets and their host star over those 4 years and recording the mid-transit time for each orbit of the warm Jupiter using \citet{Daw14a}'s $N$-body transit time code. We then calculate the deviations from a best-fit linear ephemeris at each time, which define our TTVs.

We define two metrics for distinguishing detectable TTV system from non-detectable TTV systems following \citet{Maz13}. The first metric is the ratio of the scatter in the TTVs, $s_{\rm{TTV}}$, defined as the median absolute deviation (MAD) of the TTV series, to their median error, $\bar {\sigma}_{\rm{TT}}$. 
We randomly draw $\bar{\sigma}_{\rm{TT}}$ from the median uncertainties in mid-transit time of \citet{Hol16}'s light curve fits to warm Jupiter \textit{Kepler} candidates. The second metric is the false alarm probability (FAP) of the Lomb-Scargle (LS) periodogram of the TTV time series. To generate the LS periodogram, we use the \textsc{rvlin} package \citep{Wri09}, assigning the median uncertainty to each data point in our simulated TTV time series and applying a Gaussian scatter. We categorize perturbers as detectable via TTVs when the $s_{\rm{TTV}}/\bar {\sigma}_{\rm{TT}}$ ratio is greater than 5 or the LS FAP is below $3\times10^{-4}$. Note that \citet{Maz13} set the threshold for $s_{\rm{TTV}}/\bar {\sigma}_{\rm{TT}}$ at the more conservative value of 15.

\begin{figure}
\center{\includegraphics{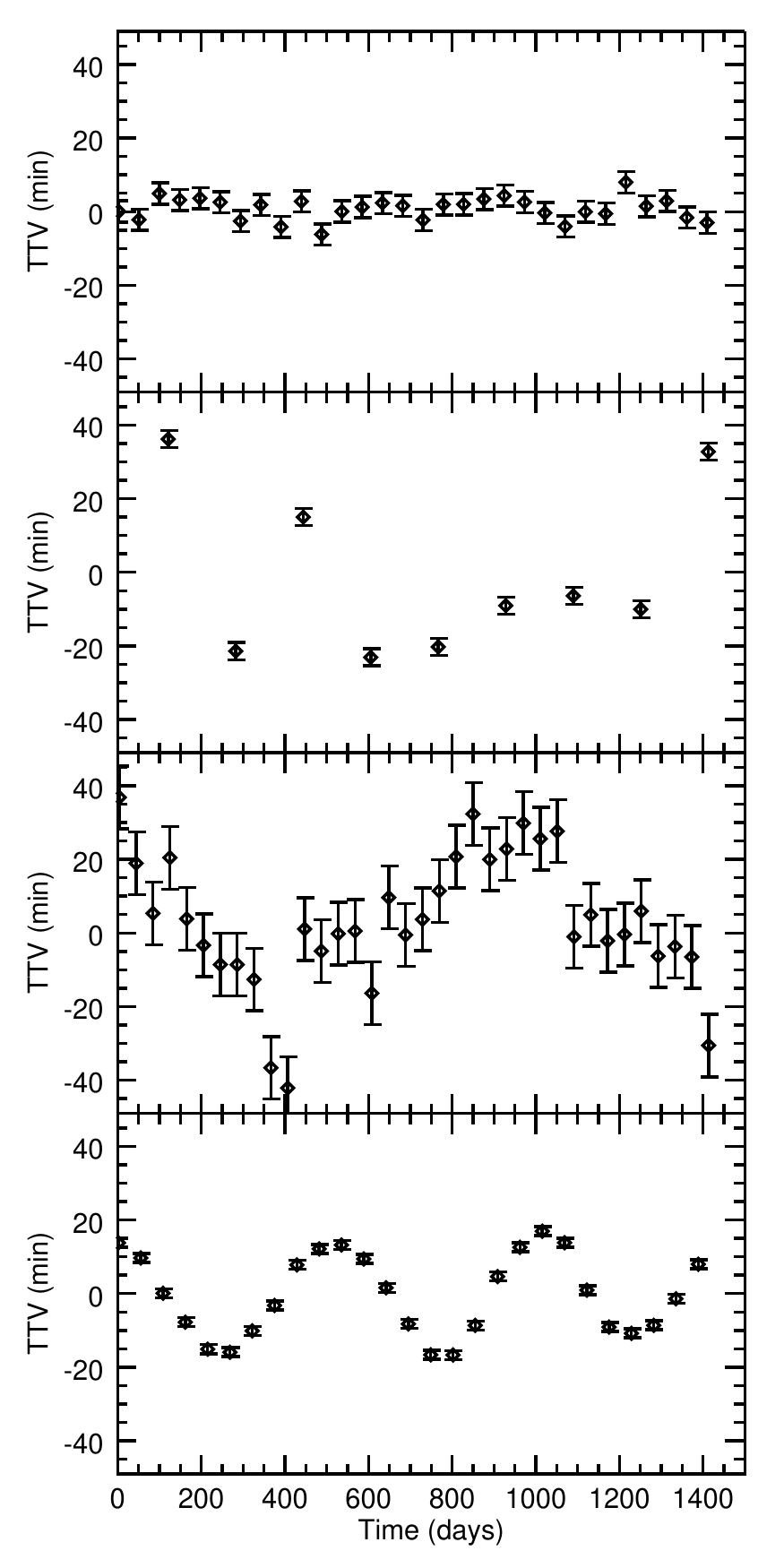}}
\caption{Four representative examples of TTV time series in our \textit{Kepler}-like population. The top panel shows the signal of an undetectable perturber ($s_{\rm{TTV}}/\bar {\sigma}_{\rm{TT}}=0.88$; LS FAP = 0.31), the second panel shows the signal of a system detectable by the ratio of scatter to median error in the TTVs ($s_{\rm{TTV}}/\bar {\sigma}_{\rm{TT}}=8.74$; LS FAP = 0.50), the third panel shows the signal of a system detectable by its periodogram signal ($s_{\rm{TTV}}/\bar {\sigma}_{\rm{TT}}=1.01$; LS FAP = $2.20\times10^{-4}$), and the bottom panel shows the signal of a perturber detectable by either metric ($s_{\rm{TTV}}/\bar {\sigma}_{\rm{TT}}=7.73$; LS FAP = $1.65\times10^{-4}$).}
\label{fig:ttv_examples}
\end{figure}

\begin{figure}
\center{\includegraphics{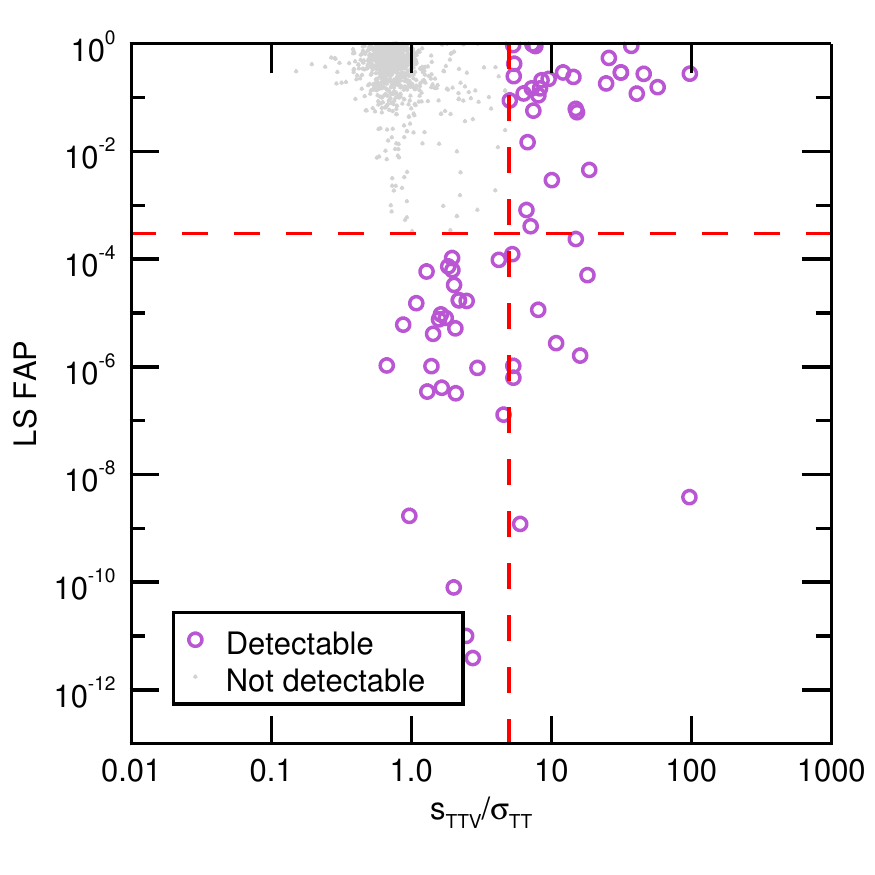}}
\caption{The majority of synthesized perturbers (92.3\%) in our \textit{Kepler}-like warm Jupiter sample induce transit timing variations (TTVs) in the warm Jupiters that are below the thresholds necessary for detection (gray points). Systems with $s_{\rm{TTV}}/\bar {\sigma}_{\rm{TT}}$ above 5 or LS FAP below $3\times10^{-4}$ are categorized as detectable (purple circles). The plot shows the value of these metrics for a random sample of 1000 systems in our \textit{Kepler}-like population.}
\label{fig:ttv}
\end{figure}

In Figure \ref{fig:ttv_examples}, we show four representative examples of TTV time series from our sample. The top panel shows the signal of an undetectable perturber, the second panel shows the signal of a perturber detectable by our first metric ($s_{\rm{TTV}}/\bar {\sigma}_{\rm{TT}}>5$), the third panel shows the signal of a perturber detectable by our second metric (LS FAP < $3\times10^{-4}$), and the bottom panel shows the signal of a perturber detectable by either metric. Each plotted time series has an uncertainty drawn from the median uncertainties in mid-transit time of \citet{Hol16}'s light curve fits to warm Jupiter \textit{Kepler} candidates and a Gaussian scatter has been applied. We present a compilation of these results from all simulated systems in Figure \ref{fig:ttv}, where purple points represent detectable systems and red dashed lines delineate our detection criteria.

Using these metrics, if each \textit{Kepler} warm Jupiter is accompanied by a perturber that fits our criteria, a small portion of these perturbers (7.7\%) should be detectable by TTVs in the \textit{Kepler} light curves. Using the more conservative cutoff of $s_{\rm{TTV}}/\bar {\sigma}_{\rm{TT}}>15$, this number drops to (4.9\%). The perturbers that are detectable are typically short period planets (<5000 day orbits) with large masses. See Section \ref{detect} for further discussion of the detectability of the perturbers. Since the majority of our synthesized perturbers produce undetectable TTVs and some TTVs have been detected for \textit{Kepler} warm Jupiters (see \citealt{Hol16}), these results do not constrain the high-eccentricity migration mechanism.

\subsection{Transit Duration Variations}\label{tdv}

Next, we calculate the transit duration variation (TDV) signal of the perturbers to the \textit{Kepler}-like warm Jupiters. We model the TDVs simultaneously with the TTVs over the 4-year \textit{Kepler} mission lifetime by tracking the slight changes to the impact parameter and transit speed of the warm Jupiter between each orbit. We calculate the approximate transit durations using the following equation from \citet{Win10b} which includes a correction for non-circular orbits:

\begin{equation}
    \label{eq:tdv}
    T_{tot}=\frac{P}{\pi}\arcsin\left[\frac{R_*}{a}\frac{\sqrt{(1+k)^2-b^2}}{\sin i}\right]\left(\frac{\sqrt{1-e^2}}{1+e\sin\omega}\right)
\end{equation}

where $P$ is the period, $R_*$ is the stellar radius, $a$ is the semi-major axis, $k$ is the planet-star radius ratio, $b$ is the impact parameter, $i$ is the inclination, $e$ is the eccentricity, and $\omega$ is the argument of pericenter. Our final TDV amplitudes are calculated by subtracting the minimum from the maximum duration among all of the transits.

As with the TTVs, we define two metrics for distinguishing detectable TDV systems from non-detectable TDV systems. Unlike TTVs where any linear trend is subtracted out, the slope of a TDV signal is meaningful. Therefore, our first metric is the strength of the linear trend in the TDV data, or more specifically, the ratio of the estimated slope over the estimated error in the slope ($|\rm{slope}|/e_{\rm{slope}}$) following \citet{Kan19}. We randomly draw the median uncertainty in transit duration from \citet{Hol16}'s light curve fits to warm Jupiter  \textit{Kepler} candidates. The second metric is the false alarm probability (FAP) of the Lomb-Scargle (LS) periodogram of the TDV time series. As with the TTVs, we generate the LS periodogram using the \textsc{rvlin} package \citep{Wri09}, assigning the median error to each data point and applying a Gaussian scatter. We categorize perturbers as detectable via TDVs when the $|\rm{slope}|/e_{\rm{slope}}$ is greater than 3.5 or the LS FAP is below $3\times10^{-4}$ \citep{Kan19}.

\begin{figure}
\center{\includegraphics{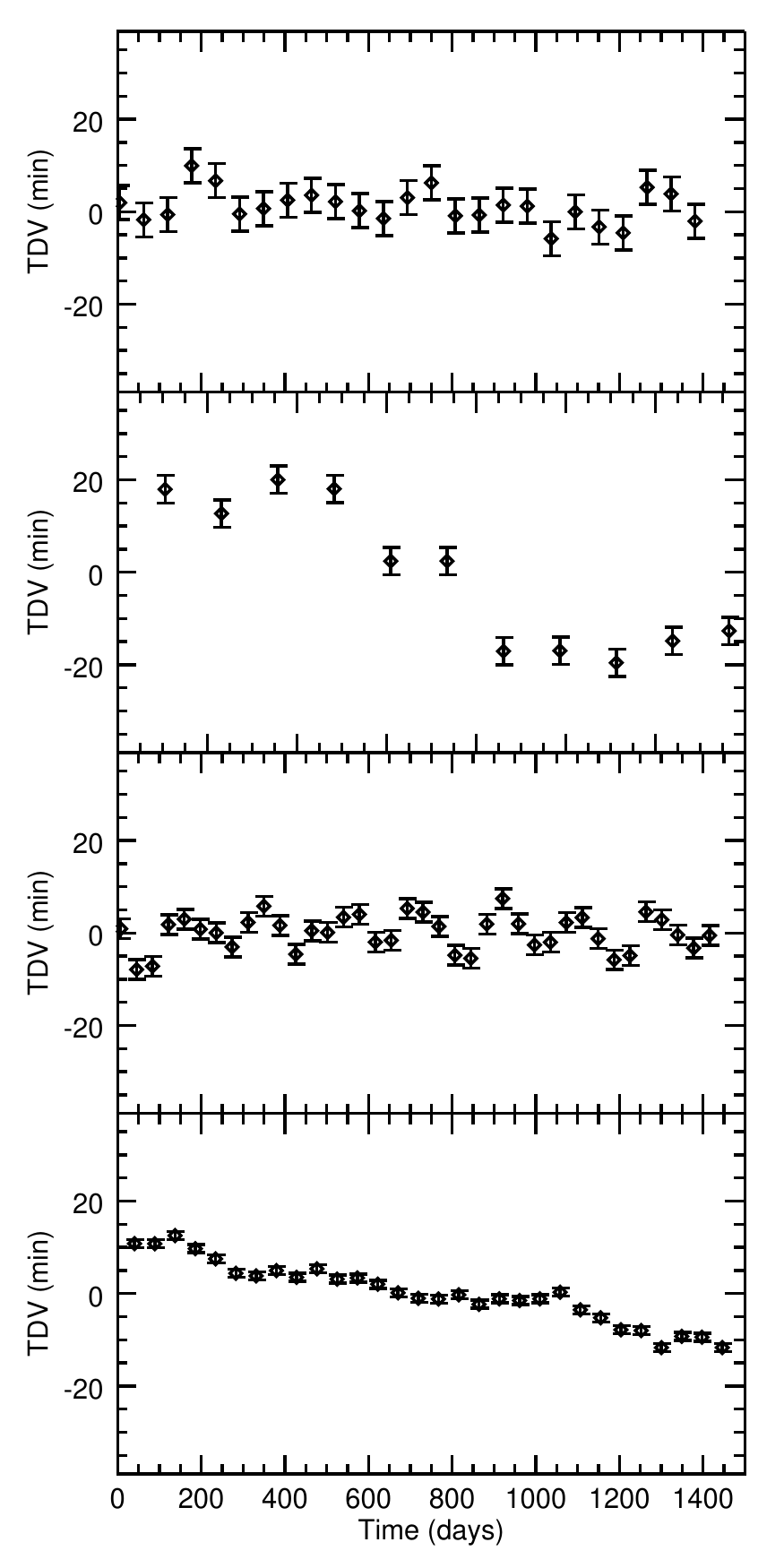}}
\caption{Four representative examples of TDV time series in our \textit{Kepler}-like population. The top panel shows the signal of an undetectable perturber ($|\rm{slope}|/e_{\rm{slope}}=1.44$; LS FAP = 0.32), the second panel shows the signal of a system detectable by the ratio of its TDV slope to the error in slope ($|\rm{slope}|/e_{\rm{slope}}=15.15$; LS FAP = 0.06), the third panel shows the signal of a system detectable by its periodogram signal ($|\rm{slope}|/e_{\rm{slope}}=0.64$; LS FAP = $5.07\times10^{-5}$), and the bottom panel shows the signal of a perturber detectable by either metric ($|\rm{slope}|/e_{\rm{slope}}=39.11$; LS FAP = $8.29\times10^{-5}$).}
\label{fig:tdv_examples}
\end{figure}

\begin{figure}
\center{\includegraphics{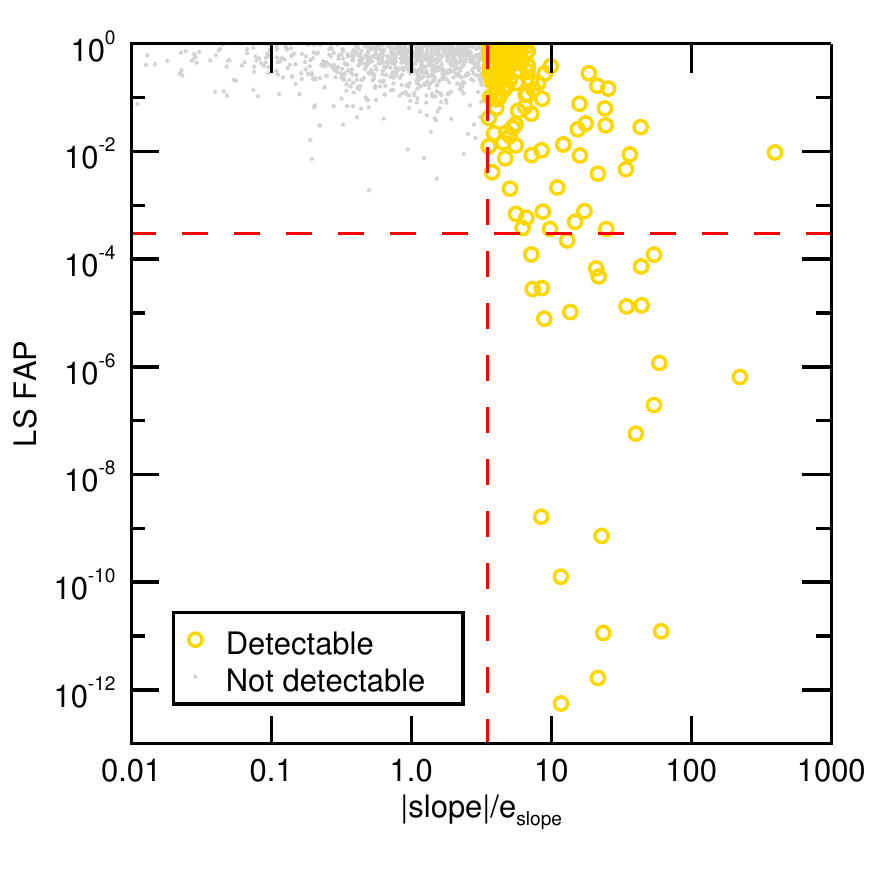}}
\caption{Same as Figure \ref{fig:ttv} but for our transit duration variation (TDV) calculation. The majority of synthesized perturbers (86.5\%) in our \textit{Kepler}-like warm Jupiter sample induce TDVs in the warm Jupiters that are below the thresholds necessary for detection (gray points). Systems with $|\rm{slope}|/e_{\rm{slope}}$ above 3.5 or LS FAP below $3\times10^{-4}$ are categorized as detectable (yellow circles). The plot shows the value of these metrics for a random sample of 1000 systems in our \textit{Kepler}-like population.}
\label{fig:tdv}
\end{figure}

In Figure \ref{fig:tdv_examples}, we show four representative examples of TDV time series from our sample. The top panel shows the signal of an undetectable perturber, the second panel shows the signal of a perturber detectable by our first metric ($|\rm{slope}|/e_{\rm{slope}}>3.5$), the third panel shows the signal of a perturber detectable by our second metric (LS FAP < $3\times10^{-4}$), and the bottom panel shows the signal of a perturber detectable by either metric. Each plotted time series has an uncertainty drawn from the median uncertainties in transit duration from \citet{Hol16}'s light curve fits to warm Jupiter \textit{Kepler} candidates and a Gaussian scatter has been applied. We present a compilation of these results from all simulated systems in Figure \ref{fig:tdv}, where yellow points represent detectable systems and red dashed lines delineate our detection criteria.

If each \textit{Kepler} warm Jupiter is accompanied by a perturber that fits our criteria from Section \ref{perturbers}, a small percentage of these perturbers (13.5\%) should be detectable by TDVs in the \textit{Kepler} light curves. Like with the TTVs, short-period perturbers with large masses are the easiest to detect. See Section \ref{detect} for further discussion of the TDV detectability of the perturbers. As with the TTVs (Section \ref{ttv}), these results alone cannot significantly constrain the high-eccentricity migration mechanism.

\subsection{Radial Velocity Trends}\label{rv}
Next, we calculate the radial velocity (RV) signal of the synthesized perturbers to our \textit{TESS}-like warm Jupiters and our RV-detected warm Jupiters. Since RV followup of known planetary systems does not come from a single instrument or survey, we model the signal with two representative time baselines (3 months and 3 years). The periods of our perturbers span the range of 200 to 100,000 days, so some signals may appear weak over a short time baseline, but strong once a larger percentage of their orbit is observed.

To calculate the observed $\Delta$RV over a given time baseline, we first model the radial motion of the star due to both planets. For the 3-month time baseline, we assume 20 data points and for the 3-year baseline, we assume 100 data points. For each point, we assume a 1 m/s uncertainty, which could arise from instrument error and/or stellar variability. Once we have our RV signal, we fit a 1-planet model to it using \textsc{mpfit} \citep{Mar09}, fixing the period and mean longitude $\lambda$ for the \textit{TESS}-like population. Our final calculated RV is the 2-planet signal with the best-fit 1-planet solution subtracted out. This approach is more conservative than simply calculating the outer planet's signal because, for a real system, we do not precisely know all of the inner planet parameters a priori and they might be conflated in the two-planet fit.

In order to determine which signals are detectable, we calculate the ratio of the slope of the RV signal to the error in the slope ($|\rm{slope}|/e_{\rm{slope}}$). As with the TDV signals, we categorize perturbers for which this metric is greater than 3.5 as detectable since those systems would have RV trends that are statistically distinguishable from zero. We also calculate a Lomb-Scargle (LS) periodogram of the RV time series, but it is ineffective at discovering the long-period signals of our perturbers, so we omit it here. 

\begin{figure}
\center{\includegraphics{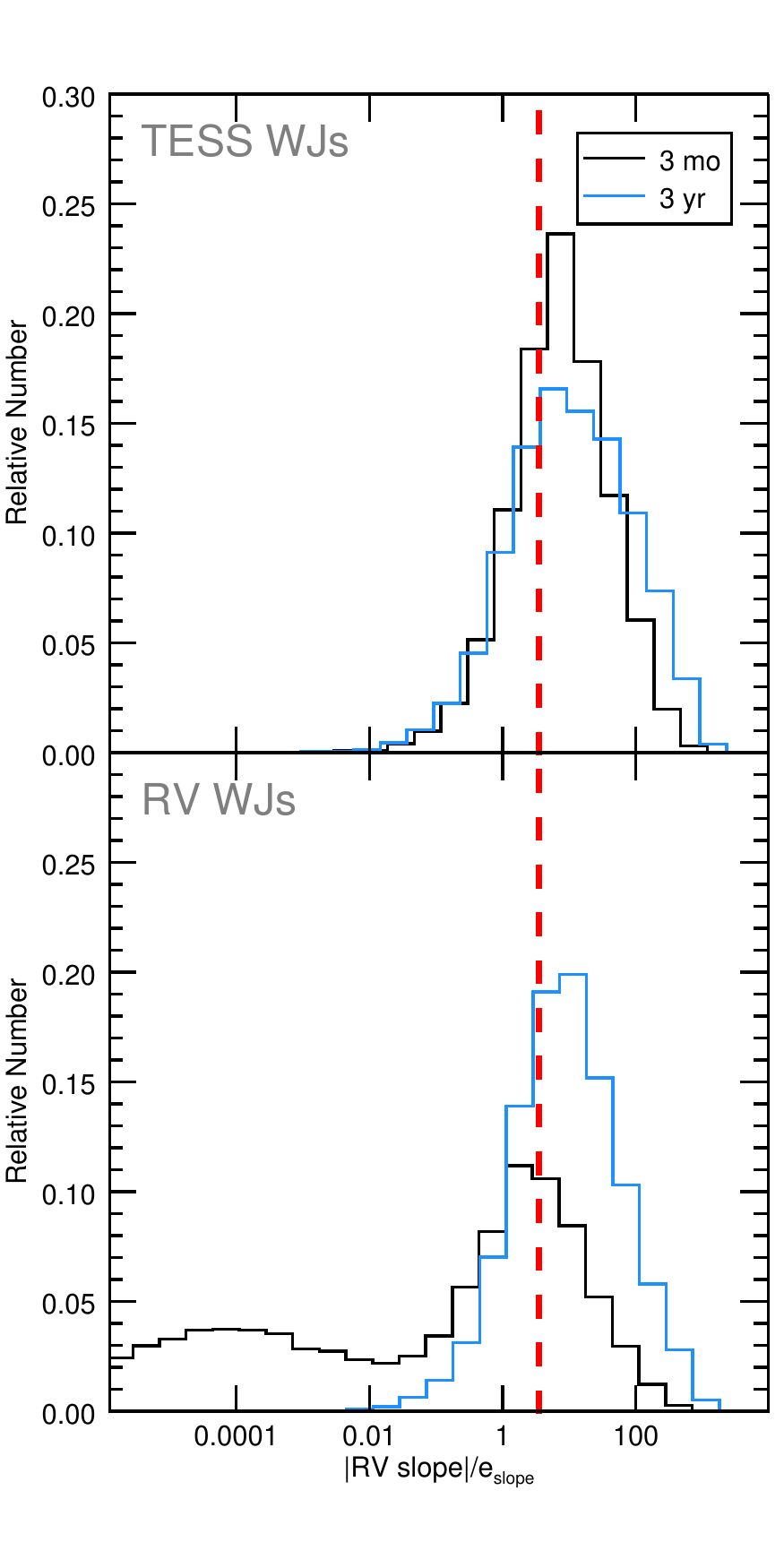}}
\caption{A large fraction of synthesized perturbers in our \textit{TESS}-like (top) and RV-detected (bottom) warm Jupiter systems are detectable via their RV signal. Here we show histograms of the $|\rm{slope}|/e_{\rm{slope}}$ metric for two representative time baselines: 3 months (black) and 3 years (blue). For each system, we calculate the 2-planet RV signal and subtract off a 1-planet warm Jupiter fit to derive at an isolated companion signal. The red dashed line is placed at 3.5, above which we describe the RV trend as detectable.}
\label{fig:rv}
\end{figure}

The results of both the 3-month calculation and the 3-year calculation are plotted in Figure \ref{fig:rv}. We find that 77.2(91.2)\% of our companions to \textit{TESS}-like warm Jupiters and 31.4(83.0)\% of our companions to RV-detected warm Jupiters would produce detectable signals over 3 months(3 years) of RV data. The bump near 0.0001 in the slope ratio for the 3-month RV-detected warm Jupiter companions is caused by degeneracies between the signals of the inner and outer planets that are broken with a longer time baseline. Note that $e_{\rm{slope}}$ is strongly dependent on the stellar activity of the host star and our assumed RV uncertainty of 1 m/s is representative of state-of-the-art RV instruments observing quiet stars. For noisier stars or less precise instruments, the detectability of our perturbers decreases significantly (e.g., our \textit{TESS} perturbers are 64.4\% detectable over 3 months of observations at 3 m/s uncertainty or 32.9\% detectable at 10 m/s uncertainty). The perturbers that are detectable by their RV slope cover the full mass range tested, with larger masses allowing for longer period detections. See Section \ref{detect} for further discussion of the detectability of the perturbers and Section \ref{companions} for a discussion of the known companions to RV-detected warm Jupiters. 

\subsection{Astrometry}\label{astro}
Finally, we calculate the astrometric signal of the synthesized perturbers to our \textit{TESS}-like warm Jupiters and our RV-detected warm Jupiters as they would be observed by \textit{Gaia} \citep{Lin96,Per01}. As with the RV calculation, we leave out \textit{Kepler} warm Jupiters here because their host stars are typically too far away for \textit{Gaia} to successfully detect an astrometric planetary signal. Specifically, \textit{Gaia} is expected to detect planets out to 500 pc \citep{Per14}, while nearly all Kepler warm Jupiter hosts are outside of that radius. We model the transverse motion of the stellar host on the sky due to the gravitational pull of the perturber, then subtract off a linear fit, which would be removed as proper motion. The following is our model for the star's motion in right ascension and declination

\begin{equation}
\label{eq:xi}
    \xi(t)=\alpha_0^*+\left(t-t_0\right)\mu_{\alpha^*}+BX(t)+GY(t)
\end{equation}

\begin{equation}
\label{eq:eta}
    \eta(t)=\delta_0+\left(t-t_0\right)\mu_{\delta}+AX(t)+FY(t),
\end{equation}

where $\alpha^*=\alpha$ cos $\delta$, $(\alpha_0^*,\delta_0)$ is the reference position of the star, $(\mu_{\alpha^*},\mu_{\delta})$ is the proper motion vector, $A$, $B$, $F$, and $G$ are the Thiele-Innes constants, and (X(t),Y(t)) is the vector describing the motion of the star in its orbit as a function of the eccentric anomaly, E. We calculate E iteratively from the mean anomaly, M, and the eccentricity, e, following \citet{Dan88}. For a detailed description of the astrometric model, see \citet{Qui10}.

\begin{figure}
\center{\includegraphics{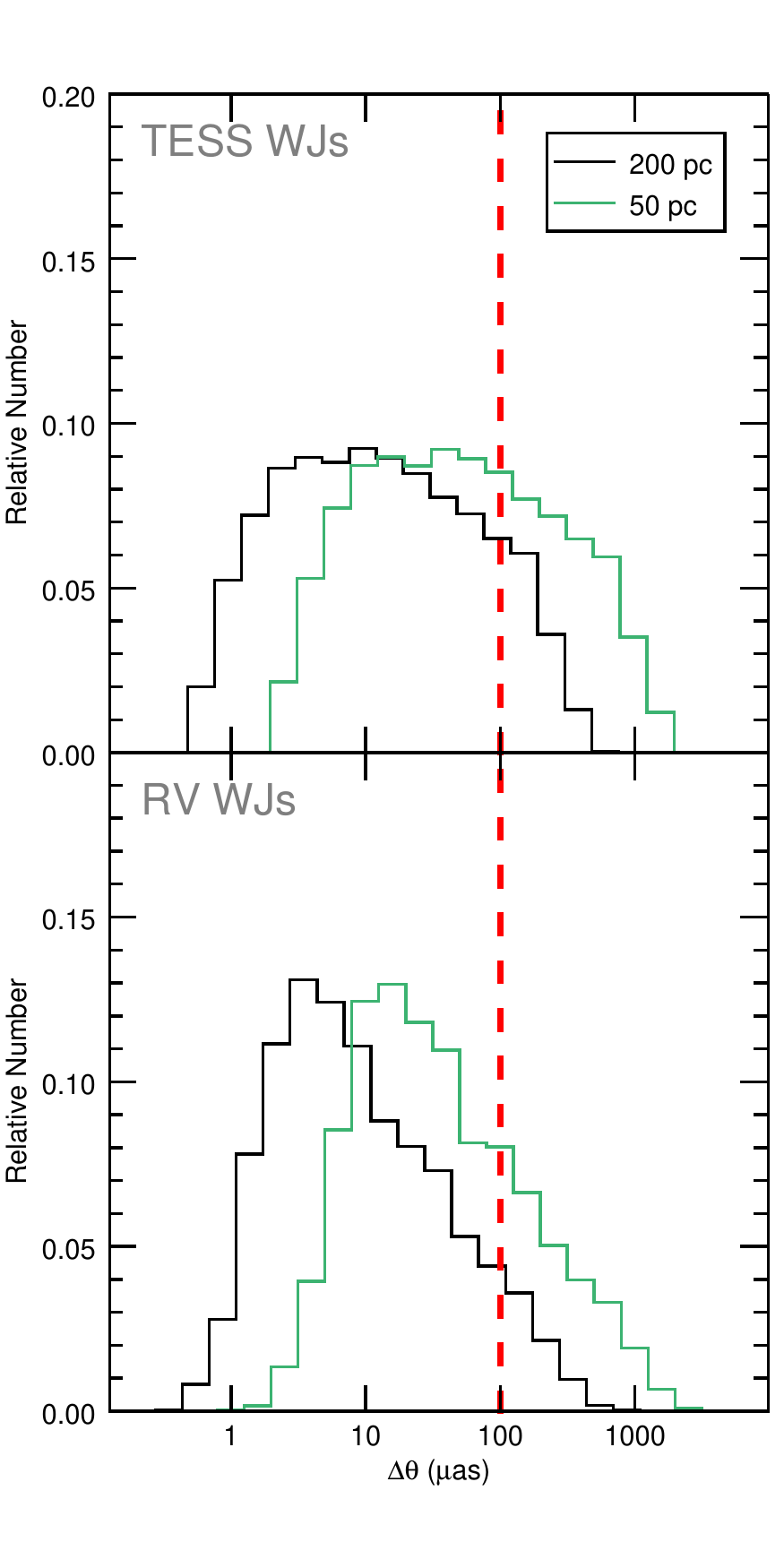}}
\caption{Same as Figure \ref{fig:ttv} but for the astrometric signal of our perturbers to \textit{TESS} warm Jupiters (top) and RV-discovered warm Jupiters (bottom). The black histogram represents the astrometric signal assuming a 200 pc distance (typical of \textit{TESS} warm Jupiter systems) and the green histogram represents the same calculation with a 50 pc distance (typical of RV warm Jupiter systems). For each system, we calculate the isolated astrometric signal due to the perturber and subtract off a linear proper motion fit. The red dashed line represents the expected minimum \textit{Gaia} sensitivity of 100 $\mu$as.}
\label{fig:astro}
\end{figure}

\textit{Gaia} is expected to observe every star in its catalogue $\sim$70 times over the course of its 5 year primary mission lifetime \citep{Jor08}. For representative distances that are consistent with typical \textit{TESS} warm Jupiters (200 parsecs) and typical RV Warm Jupiters (50 parsecs), all of which will appear in the \textit{Gaia} astrometric catalogue, we calculate the maximum angular distance, $\Delta\theta$, between data points in the model for each of our simulated systems after subtracting off the linear proper motion fit. We plot the results of this calculation for \textit{TESS}-like warm Jupiter systems and RV-detected warm Jupiter systems in Figure \ref{fig:astro}. Also plotted in Figure \ref{fig:astro} is a line indicating the conservative \textit{Gaia} detection limit for stars brighter than $G=12$ mag of $\sim$100 $\mu$as \citep{Per14}. 16.7(40.3)\% of the perturbers in \textit{TESS}-like warm Jupiter systems and 9.8(29.6)\% of the perturbers in RV-discovered warm Jupiter systems would be detectable with Gaia astrometry at distances of 200(50) parsecs. Only perturbers with large masses (>1 $M_{\rm{Jup}}$) are detectable with this method and intermediate periods are preferred. See Section \ref{detect} for further discussion of the detectability of the perturbers.

\subsection{Detectability} \label{detect}
As shown in sections \ref{ttv} and \ref{tdv}, a small portion of our perturbers should be detectable in the current \textit{Kepler} light curves via TTVs or TDVs (19.0\%, combined). However, because most of the \textit{Kepler} stars are too dim and too far away for RV or astrometric planet detection, most of the perturbers in those systems will remain hidden, even if they exist. On the other hand, many of the perturbers in \textit{TESS} and RV-discovered warm Jupiter systems (77.2\% and 31.4\%, respectively) are observable with high precision RV followup with even a short time baseline (see Section \ref{rv}). \textit{Gaia} will also be able to detect many perturbers associated with \textit{TESS} and RV-discovered warm Jupiters (16.7\% and 29.6\%, respectively; see Section \ref{astro}). Furthermore, when combining the RV and astrometric detection methods together, nearly all of the parameter space for perturbers to \textit{TESS} warm Jupiter and most of the parameter space for RV-discovered warm Jupiters that satisfy the two constraints in Section \ref{perturbers} are detectable with current or near future instruments (78.3\% and 45.8\%, respectively).

\begin{figure*}
\center{\includegraphics{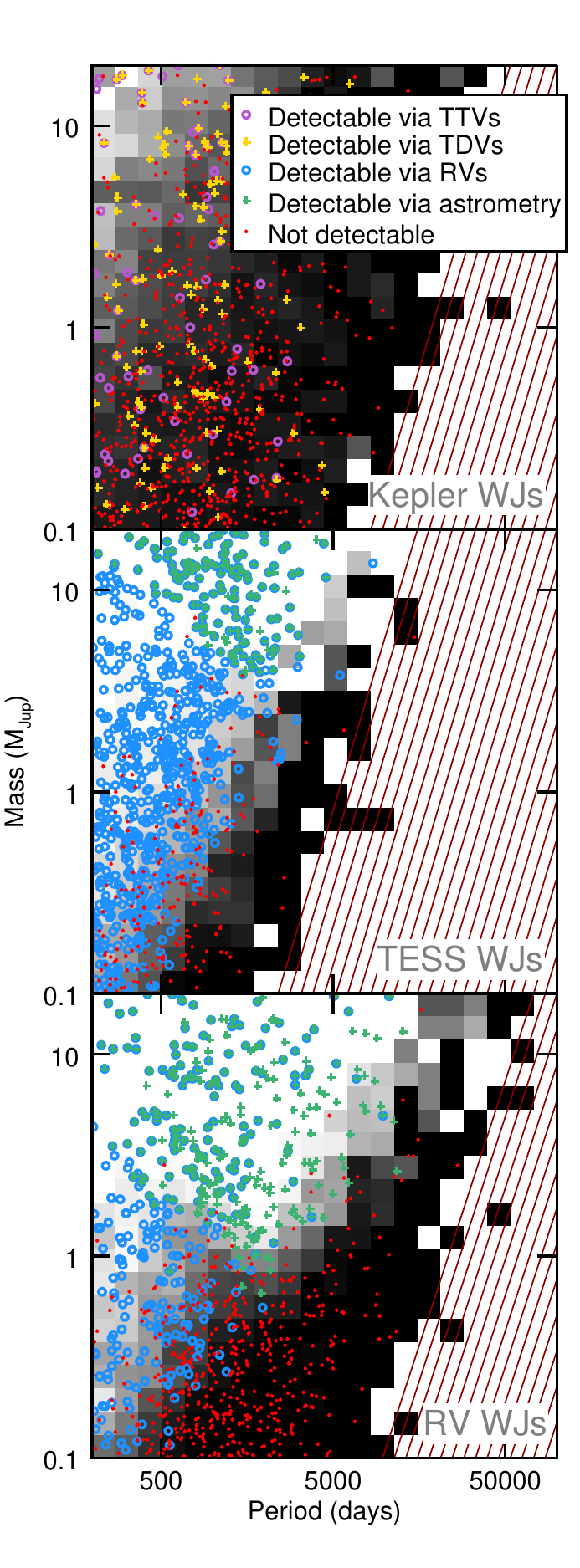}}
\caption{The majority of our simulated  perturbers to \textit{TESS} (middle) and RV-discovered (bottom) warm Jupiters are detectable by their radial velocity and astrometric signal, while only a small percentage of our simulated perturbers to \textit{Kepler} (top) warm Jupiters are detectable by their transit timing and duration variations. Here we show the mass vs. period of these perturbers, as these both have strong effects on the signal of all four observables. In each panel, we plot a random sample of only 1000 systems for clarity. The black shading is a contour map showing the ratio of detectable perturbers to undetectable perturbers such that whiter areas of parameter space imply more detections. The lower right area of parameter space (red shaded area) is largely undetectable, but there are no perturbers in those regions that satisfy our perturber strength criteria from Section \ref{perturbers}.}
\label{fig:detect}
\end{figure*}

In Figure \ref{fig:detect} we show the detectability of all of our synthetic perturbers with current and near-future instruments in terms of their masses and periods. We classify a perturber to a \textit{Kepler} warm Jupiter as detectable via its transit timing variation (TTV) signal if $s_{\rm{TTV}}/\bar {\sigma}_{\rm{TT}}>5$ or LS FAP$<3\times10^{-4}$ (Purple circles). We classify a perturber to a \textit{Kepler} warm Jupiter as detectable via its transit duration variation (TDV) signal if $|\rm{slope}|/e_{\rm{slope}}>3.5$ or LS FAP$<3\times10^{-4}$ (yellow crosses). We classify a perturber to a \textit{TESS} or RV-discovered warm Jupiter as detectable via its observable radial velocity (RV) trend if $|\rm{slope}|/e_{\rm{slope}}>3.5$ over 3 months of observations (blue circles). Lastly, we classify a perturber to a \textit{TESS} or RV-discovered warm Jupiter as detectable via its astrometric signal ($\Delta\theta$) if $\Delta\theta>100$ $\mu$as assuming a distance of 200 parsecs for \textit{TESS} systems and 50 parsecs for RV systems (green crosses). Red points represent undetectable companions in each sample. The background shading demonstrates the percentage of perturbers that are detectable in a given grid space, with whiter areas being more detectable. The lower-right corner in each panel is an area of parameter space that did not produce any perturbers capable of passing the criteria from Section \ref{perturbers}. The three panels in Figure \ref{fig:detect} show the detectability of perturbers in our \textit{Kepler}-like (top), \textit{TESS}-like (middle), and RV-discovered (bottom) samples. As stated above, 19.0\% of our perturbers to \textit{Kepler} warm Jupiters should currently be detectable in the \textit{Kepler} light curves and 78.3(45.8)\% of our perturbers to \textit{TESS}(RV-discovered) warm Jupiters will be detectable in the near future with RV and astrometric followup. Those that are undetectable tend to have low masses and intermediate-to-long periods, as these areas of parameter space are the most difficult to detect with RVs and astrometry and tend to produce weaker TTV and TDV signals.

If, after sufficient RV followup of \textit{TESS} warm Jupiters and the release of the planets detected by \textit{Gaia}, we do not find a large number of strong perturbers accompanying warm Jupiters, we can infer that perturber-coupled high-eccentricity migration may not be a common mechanism for delivering warm Jupiters.

\section{Testing our assumed parameter distributions}\label{pop_comp}
Here we consider alternative choices for warm Jupiter and perturber population parameters that are not well-constrained observationally. Our fiducial warm Jupiter populations are described in Sections \ref{kepler_warm}, \ref{tess_warm}, and \ref{rv_warm} and consist of gas giant planets with periods between 10 and 200 days and a wide range of eccentricities. We examine these parameter ranges in Section \ref{wj_param}. Our fiducial perturber population is described in Section \ref{perturbers} and includes masses between 0.1 and 20 $m_{\rm{Jup}}$ drawn from the \citet{Cum08} power law fit, periods drawn from the \citet{Fer19} broken power law fit,
eccentricities drawn from a beta distribution fit, sky plane inclinations drawn from an isotropic distribution, and all other orbital angles drawn from uniform distributions between 0 and $2\pi$. We test these parameter distributions in Sections \ref{inclination}, \ref{eccentricity}, \ref{period}, and \ref{mass}. Because the fraction of \textit{Kepler}-like perturbers detectable by their transit timing and duration variations is small, we focus on only the \textit{TESS}-like and RV-discovered systems in these sections. We note, however, that when we apply these same tests to the \textit{Kepler}-like population, the fraction of detectable perturbers remains near $\sim20\%$ in all cases.

Figure \ref{fig:comparison} shows the calculated RV and astrometry signals for all of the perturber parameter distributions we test in Sections \ref{inclination}, \ref{eccentricity}, and \ref{period}. We assume a 3 month time baseline for the RV calculations. For the astrometric calculations, we assume a distance of 200 pc for the \textit{TESS}-like warm Jupiters and 50 pc for the RV-discovered systems. The details of each alternative parameter distribution are discussed in the following subsections, but Figure \ref{fig:comparison} generally shows that our results are insensitive to the particular perturber population used for the calculations. We test the sensitivity of our results to changes in the inclination (Section \ref{inclination}), eccentricity (Section \ref{eccentricity}), and period (Section \ref{period}) distributions. For all of these tests, we maintain the power law mass distribution fit from \citet{Cum08}. Although the mass and period distributions are dependent on one another, \citet{Bry16} and \citet{Fer19} both find consistency with the \citet{Cum08} mass distribution, but inconsistency with their period distribution. We test the impact of the range of masses we explored in Section \ref{mass}.

\subsection{Warm Jupiter Parameter Ranges}\label{wj_param}
Thus far in this study, we have considered perturbers to warm Jupiters from the full range of periods (10-200 days). Here, we focus only on short-period warm Jupiters ($P<50$ days) and rerun our calculations in order to determine if a more focused future search might produce more robust results. We find that this period cut slightly shrinks the allowed area of the parameter space for perturbers, which, in turn, increases the detectability of the perturbers. This effect is very small for the TTV, TDV, and astrometric detection techniques as well as for the RV-detected perturbers to \textit{TESS} warm Jupiters which tend to have short periods anyway. However, The RV-detectability of RV-discovered warm Jupiters is approximately double that of our prior results (62.7\% over a 3-month baseline).

Until this point, we have also assumed all warm Jupiters may have been delivered by the high-eccentricity tidal migration mechanism, even those with low observed eccentricities. However, it may be likely that most low eccentricity warm Jupiters arrived via a different formation or migration method (e.g., \citealt{Daw13,Pet16}). To test the effects of our choice on the results of this paper, we rerun our calculations considering only perturbers to warm Jupiters with large eccentricities ($e>.4$). We find that the detectability of these perturbers is not significantly larger than our prior result for any of the considered detection techniques.

\subsection{Perturber Mutual Inclination}\label{inclination}
We now examine the effects of the inclination distribution of our perturbers. Because the true inclination distribution of giant planet companions is not well-constrained, we used an isotropic distribution for our fiducial population. Here we compare this to the other extreme: coplanar companions. In the coplanar case, we assign the inclinations ($i$) and longitudes of ascending node ($\Omega$) of the perturbers to be identical to those of the warm Jupiter. We find the observed RV trends for both populations of warm Jupiters to be indistinguishable between the coplanar case and fiducial isotropic case. The astrometric signal of the coplanar perturbers is shifted towards smaller $\Delta\theta$ in the transiting warm Jupiter case. This can be attributed to the fact that, in edge-on systems, one dimension of astrometric movement is lost. In the RV-discovered warm Jupiter case, this discrepancy is not present because the sky-plane inclinations of these systems are widely distributed. In general, our assumption of isotropic perturbers has a small effect on our results, but may predict more astrometric detections for \textit{TESS} companions than we would expect from a coplanar inclination distribution.

\begin{figure*}
\includegraphics{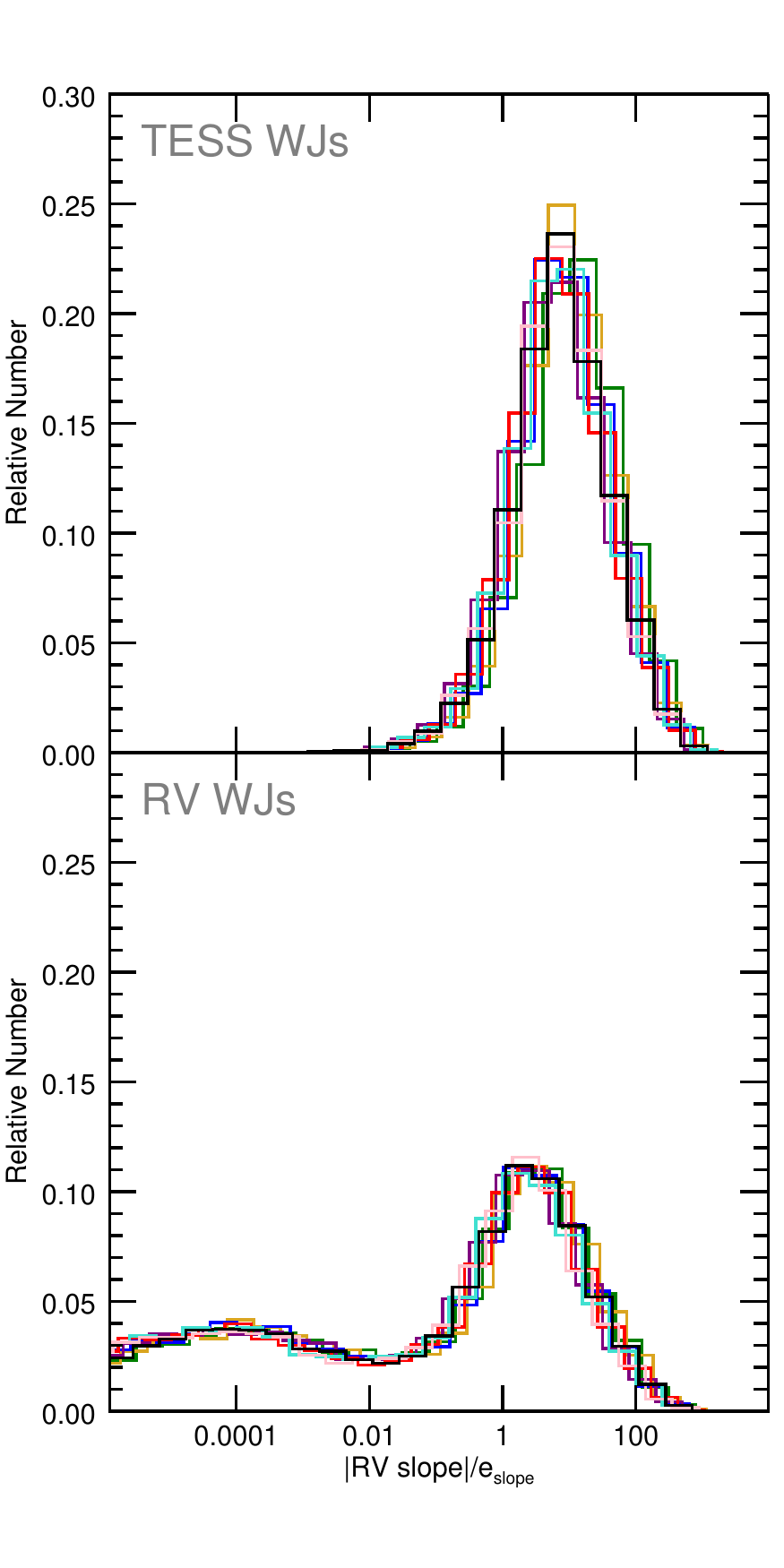}
\includegraphics{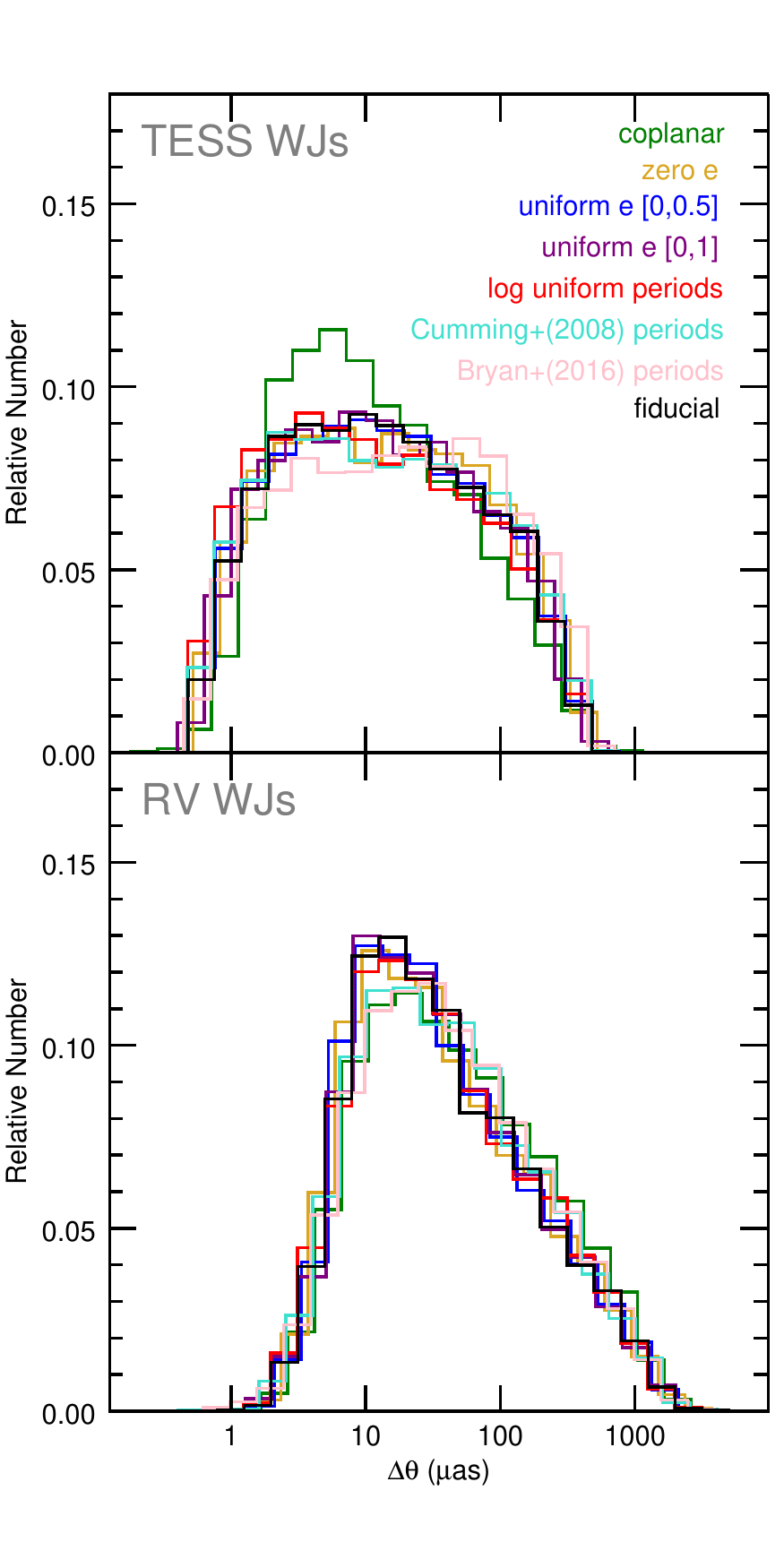}
\caption{The results of our calculated observables are largely insensitive to changes in the underlying perturber parameter distributions. Each histogram on this plot shows the recalculated RV and astrometric signals for our perturbers assuming different underlying distributions for their period, eccentricity, and inclination, as described in Section \ref{pop_comp}.}
\label{fig:comparison}
\end{figure*}

\subsection{Perturber Eccentricity}\label{eccentricity}
We now examine the eccentricity distribution of our perturbers. Our fiducial perturber population drew eccentricities from a Beta distribution with $\alpha=0.74$ and $\beta=1.61$ (see Section \ref{perturbers}). In comparison to this, we first test a uniform eccentricity distribution between $e$=0-0.5. While a maximum $e$ of 0.5 is not strongly observationally motivated, it is consistent with assumptions used by previous theoretical studies on this topic \citep{Don14,Pet16}. Additionally, we test the two extreme cases of zero eccentricity and uniform eccentricities between $e$=0-1, the latter of which which is more representative of the observed binary stellar companion sample \citep{Duc13}.

We find that, even including the extreme cases, all of our eccentricity distributions produce similar distributions of observables for each case we test. As shown in Figure \ref{fig:comparison}, each eccentricity test closely follows the fiducial population histogram.

\begin{deluxetable*}{llll}
\tablecaption{Fitting parameters for each tested period distribution.\label{tab:periods}}
\tablehead{
\colhead{Distribution} & \colhead{Parameter 1} & \colhead{Parameter 2} & \colhead{Reference}}
\startdata
log uniform() & - & - & - \\[1pt]
power law($\alpha$) & 0.26 & - & \cite{Cum08} \\[1pt]
power law($\alpha$) & 0.38 & - & \cite{Bry16} \\[1pt]
broken power law($P_1$,$P_{\textrm{break}}$) & 0.63 & 859 days & \cite{Fer19} \\[1pt]
\enddata
\end{deluxetable*}

\subsection{Perturber Orbital Period}\label{period}
Next, we consider how the underlying perturber orbital period distribution we draw from affects our calculated observables. Our fiducial periods are drawn from a broken power law distribution. We note that our perturber periods, masses, and eccentricities are drawn before applying the two analytical cuts described in Section \ref{perturbers}. Since any perturbers that do not satisfy the stability or perturber strength criteria are discarded and redrawn, the final distributions are different from the underlying distributions we draw from. The filtering by these criteria is particularly important for orbital periods, where short and long periods are disfavored by the two cuts and intermediate periods are selectively chosen. 

\begin{figure}
\center{\includegraphics{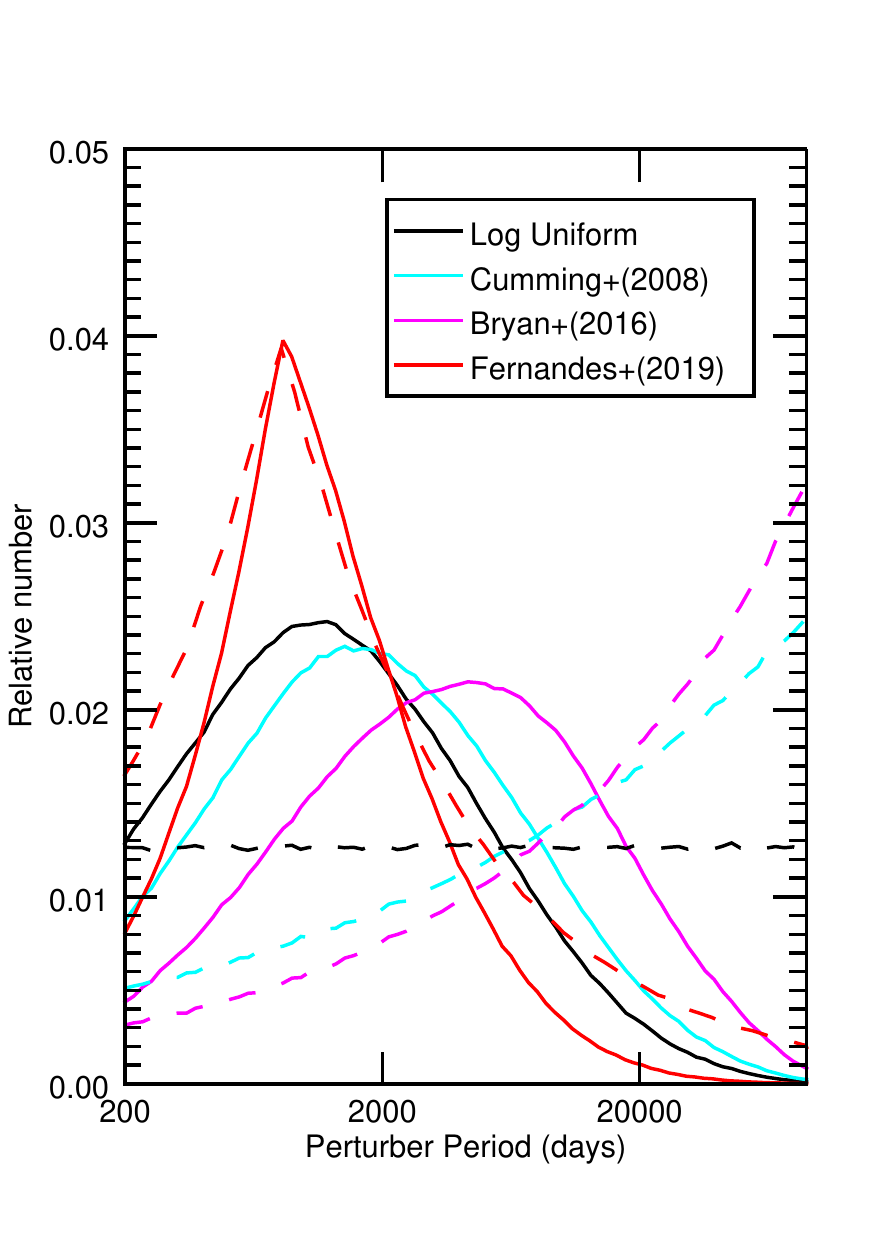}}
\caption{Perturber period distributions before (dashed line) and after (solid line) perturber strength and stability cuts (see Section \ref{perturbers}) are applied. These cuts reduce the number of short and long period perturbers and favor intermediate periods near $\sim$1000-5000 days.}
\label{fig:periods}
\end{figure}

In Figure \ref{fig:periods}, we show the four period distributions that we test, before and after applying our two cuts: a log-uniform distribution, power law fits from \citet{Cum08} and \citet{Bry16}, and a broken power law fit from \citet{Fer19}. \citet{Cum08} found evidence supporting a rising power law period distribution for giant planets, but had relatively few data points and were limited to periods under 2000 days. Using new RV data, \citet{Bry16} confirmed the rising power law trend in orbital period for short period giant planets. However, when they assessed the period distribution for companions to short period giant planets discovered by long term radial-velocity monitoring and adaptive optics (AO) imaging, they found evidence for a decreasing power law coefficient for giant planets at larger periods and speculated that a turnover in the period distribution must occur between 3 and 10 au. Here we use their single power law fit for planets with masses between 0.5 and 20 $M_{\rm{Jup}}$ and semi-major axes between 5 and 50 au, but note that the results of this fit were very sensitive to the particular mass and semi-major axis ranges chosen. With an even larger set of RV data, \citet{Fer19} modeled the turnover in the period distribution as a broken power law, which we use as our fiducial period distribution. We use the power law fits to the RV data from each study in our comparison here, whose fitting variables are recorded in Table \ref{tab:periods}.

The four different functions used to create our period distributions are significantly different in their shapes and the locations of their peaks. However, as shown in Figure \ref{fig:periods}, after accounting for perturber strength and stability, all four functions peak at intermediate periods and drop off toward longer or shorter periods. Because the final distributions after the criteria are applied in all cases are similarly shaped, the RV signal is insensitive to the particular underlying period distribution chosen. The astrometric signal is slightly affected by this choice due to its strong dependence on planet's semi-major axis. Thus, the \citet{Bry16} fit, which had the longest peak period, produces a slightly stronger astrometric signal. However, these effects are small and the \citet{Fer19} distribution, which we use in the analysis of our results, is the most conservative of the four in terms of its astrometric signal.

If all giant planets form beyond the ice line and those that we observe interior to that point are the result of migration, we might expect the companions to planets migrating via high-eccentricity migration to remain at large separations for their entire lifetime. Throughout this work, we have allowed our coupled perturbers to exist anywhere outside of a 200 day orbit. We test how this assumption affects our results by removing any perturbers interior to the ice line and rerunning our calculations. The exact location of the ice line is under investigation, but here we use 2.7 au as an illustrative value \citet{Lec06}. We find that the RV detectability of these planets is not significantly affected by this change, while the TTV/TDV detectability is decreased, and the astrometric detectability is increased. This is consistent with our previous findings that, if these perturbers exist, we do not expect to find many in the \textit{Kepler} data, but should find many by following up \textit{TESS} and RV-discovered warm Jupiters. In fact, if the perturbers are all beyond the ice line, \textit{Gaia} will be much more effective at finding them.

\subsection{Perturber Mass Range}\label{mass}
Lastly, we consider the effects of our choice for the upper limit  on the mass of our perturbers. All of the perturbers described above are of planetary mass ($<20m_{\rm{Jup}}$). Stellar perturbers could also produce the desired eccentricity excitation in the warm Jupiter. However, since binary systems harboring planets are preferentially widely separated \citep{Egg11,Moe19} and Equation \ref{eq:dong} favors smaller semi-major axes as well as larger masses, the allowed window for stellar perturbers in our parameter space is small (masses between $\sim$1 and 2 $m_{\sun}$ and semi-major axes between $\sim$100 and 150 AU). Moreover, widely separated stellar-mass perturbers generally lead to fast migration and, therefore, produce very few warm Jupiters \citep{Pet15a,Pet16,And16}. Thus, we focus on planetary companions in this paper.

We chose an upper limit on our perturber mass range of $20m_{\rm{Jup}}$ for consistency with \citet{Fer19}. Other occurrence rate studies, however, use more conservative mass cut-offs of $10m_{\rm{Jup}}$ \citep{Cum08} or $13m_{\rm{Jup}}$ \citep{Bry16}. Here we test how excluding the most massive perturbers in our study affects the detectability of the planets. When we limit the mass of the perturber to $10m_{\rm{Jup}}$, we find that the overall detectability of the perturbers is $\sim$2-4\% lower. Alternatively, if we were to include more brown dwarfs, the detectability would slightly increase.

We find that our results are largely insensitive to changes in the underlying perturber parameter distributions. For a variety of assumptions about the eccentricity, inclination, period distribution, and mass range, most perturbers that meet the criteria for strength and stability necessary for perturber-coupled high-eccentricity tidal migration of \textit{TESS} and RV-discovered warm Jupiters should be detectable within the next few years with RV and astrometric instruments.

\section{Observables of warm Jupiter perturbers from a population synthesis study}\label{petro}

Next we compute observability metrics for a population of warm Jupiter perturbers from a population synthesis study. \citet{Pet16} investigated the properties of hot and warm Jupiters produced by planet-planet Kozai-Lidov high-eccentricity tidal migration. This study assumed Sun-like host stars; a limited range of inner and outer planet semi-major axes, eccentricities, and masses; and uniformly distributed mutual inclinations. The narrow semi-major axis range allowed for migration from $\sim1$ au orbits, while preserving the dynamical stability of the system (see, e.g., \citealt{Ant16}). They evolved these systems forward in time until one of the following outcomes occurred: (1) the inner planet evolved into a hot Jupiter, (2) the inner planet was tidally disrupted, or (3) the inner planet survived for the length of the simulation (1 Gyr). 2.6\% of inner planets ended their simulations as warm Jupiters. We take the set of warm Jupiter-perturber end-states from \citet{Pet16} and calculate the observables of the perturbers following the procedure from Sections \ref{ttv}, \ref{tdv}, \ref{rv}, and \ref{astro}. The warm Jupiter-perturber systems from \citet{Pet16} are different from our earlier systems because (1) they are drawn from narrower parameter distributions and (2) they migrate in the simulations while our systems only pass a minimum analytical requirement for migration (equation \ref{eq:dong}).

\begin{figure*}
\begin{center}
\includegraphics{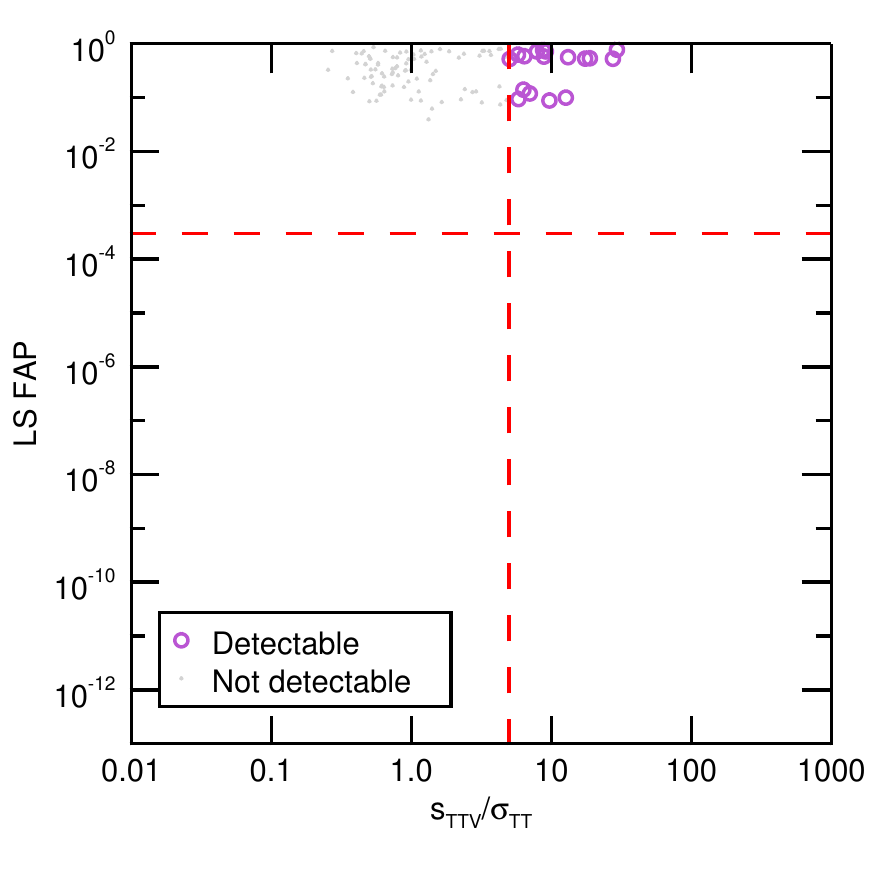} 
\includegraphics{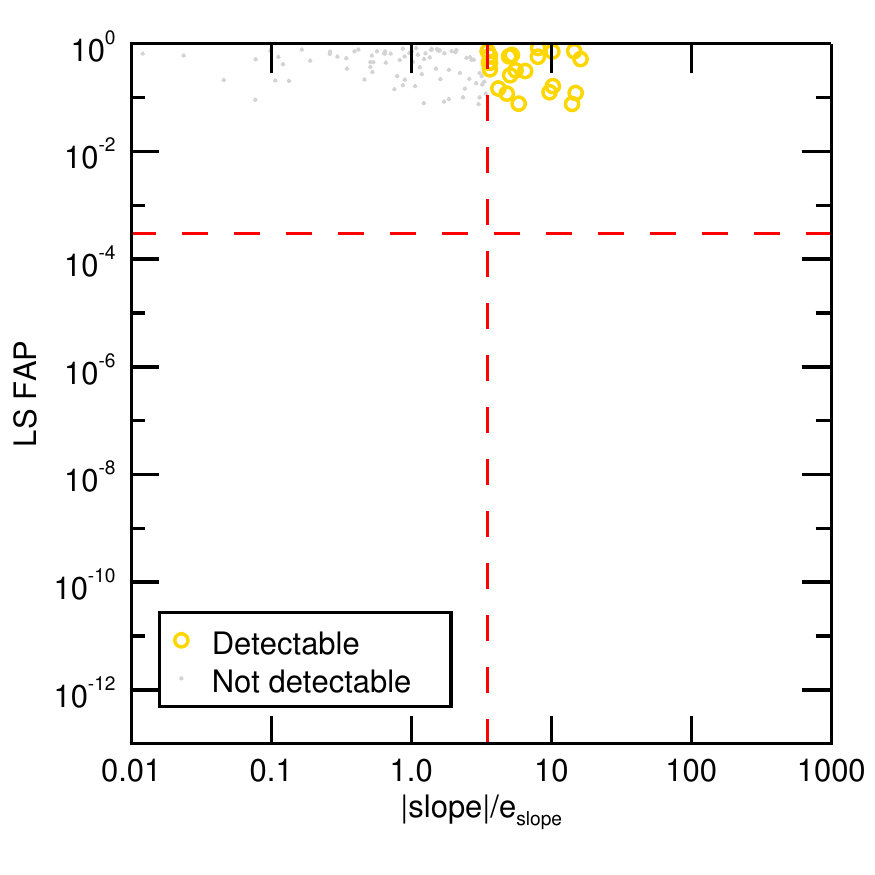}\\
\includegraphics{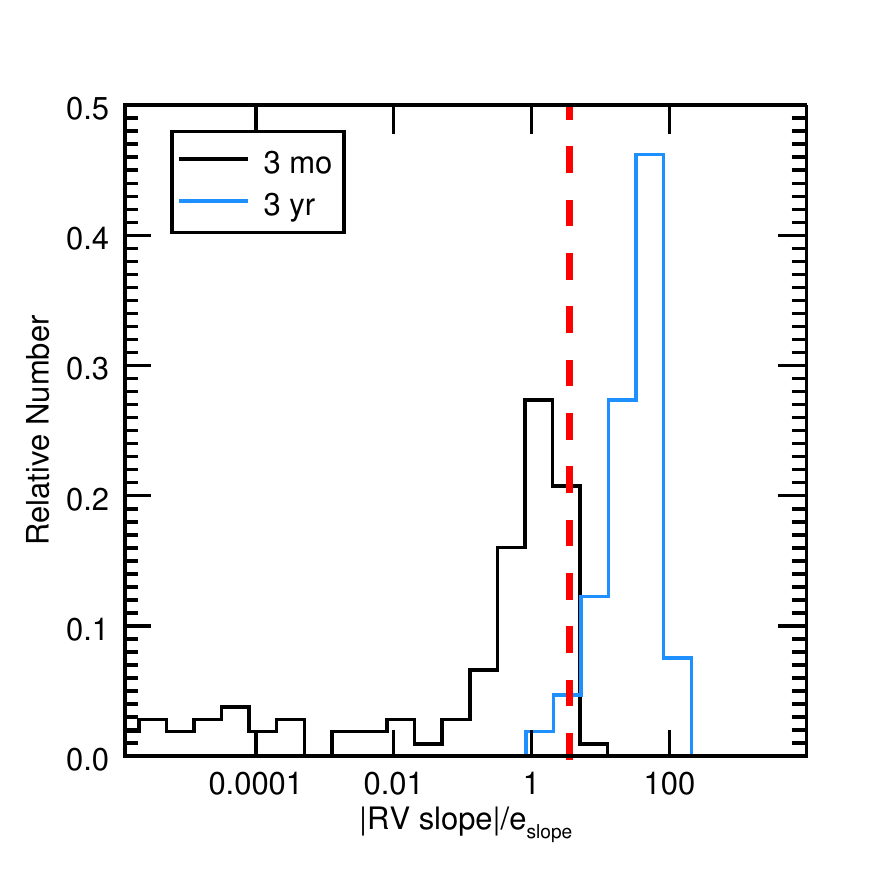}
\includegraphics{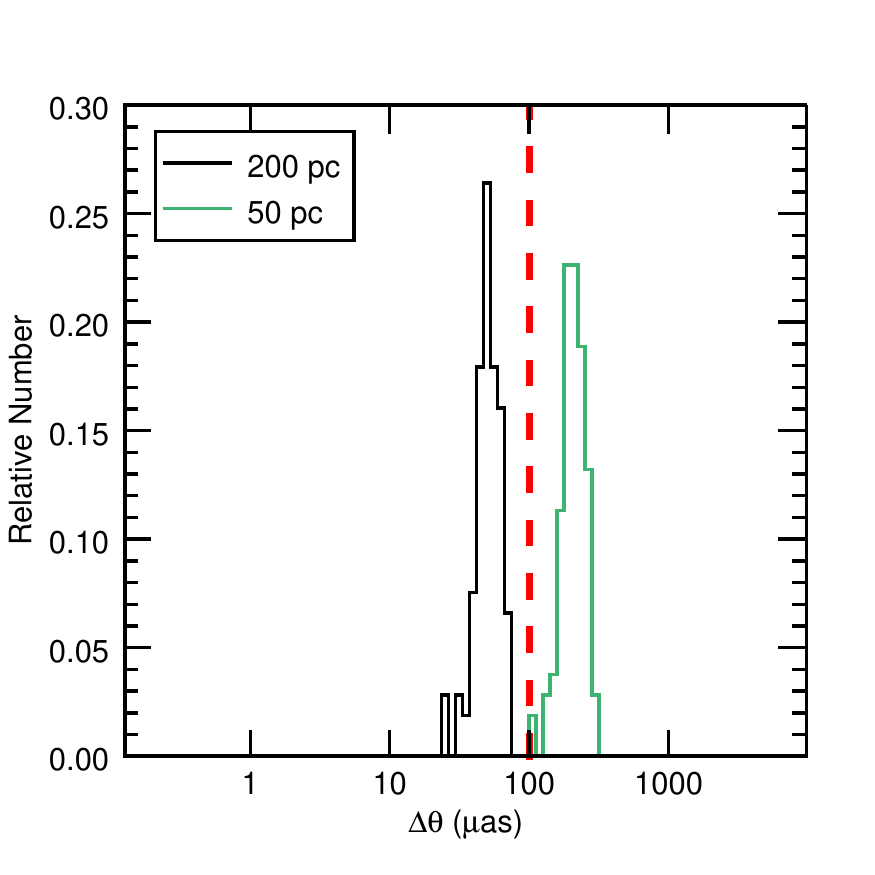} 
\end{center}
\caption{Here we show the detectability of the simulated perturbers from \citet{Pet16} based on their TTV signal ($s_{\rm{TTV}}/\bar {\sigma}_{\rm{TT}}$, periodogram false alarm probability; top left), TDV signal ($|\rm{slope}|/e_{\rm{slope}}$, periodogram false alarm probability; top right), RV signal ($|\rm{slope}|/e_{\rm{slope}}$; bottom left), and astrometric signal ($\Delta\theta$; bottom right). For the RV calculation, time baselines of 3 months (black) and 3 years (blue) are shown. For the astrometric calculation, distances of 50 pc (green) and 200 pc (black) are shown. Detection limits are shown as red dashed lines.}
\label{fig:pet}
\end{figure*}

We present the results of our TTV, TDV, RV, and astrometric calculations on this additional data set in Figure \ref{fig:pet}. We find that 33\% of these perturbers are detectable via their transit light curves (purple and yellow circles), assuming \textit{Kepler} sensitivity and time baselines. This general result is consistent with our results from Sections \ref{ttv} and \ref{tdv}. However, in this case, no planets are detectable by their periodogram signal. This result can be attributed to the fact that the end state orbital periods of the simulated warm Jupiters are consistently longer than those drawn from the power law distribution in Section \ref{pop}. Since these planets have fewer transits and, therefore, fewer points in the TTV and TDV times series, the periodogram power is weaker and no perturbers are detectable.

For both the RV and astrometric signals, the results for these perturbers occupy a narrow range, primarily due to the fact that the planets occupy narrow ranges in their orbital periods and masses. The only exception to this is that the distribution of short-baseline RV signals has an extended tail due to degeneracies with subtracting out the warm Jupiter, as we saw in Section \ref{rv}. We find that nearly all of the perturbers are detectable with 3 years of RV data (blue histogram) or by \textit{Gaia} astrometry at 50 pc (green histogram), but nearly all are undetectable with 3 months of RV data or by \textit{Gaia} astrometry at 200 pc (black histograms). If these simulated systems are more representative of the true set of systems that have engaged in perturber-coupled high-eccentricity migration than our synthesized systems, the perturbers will be more difficult to detect.

\section{Observed Warm Jupiter Companions}\label{companions}

If perturber-coupled high-eccentricity migration is a common pathway for warm Jupiter formation, the observed sample may already contain some strongly perturbing companions in warm Jupiter systems, particularly via radial velocity measurements in RV-discovered warm Jupiter systems or Kepler warm-Jupiter systems bright enough for RV followup. Since there has been no uniform survey searching for long-period planets in warm Jupiter systems, we cannot directly compare our results to the observations. However, considering the currently known companions to warm Jupiters can give us a sense of what such a survey might find. Here we compile and discuss the known massive companions to warm Jupiters. We note that \citet{Bry16} surveyed for companions to short-period planets, but did not distinguish warm Jupiters from other short-period planets they followed up.

Of the 102 RV-discovered warm Jupiters (10-200 day periods, .1-10 $m_{\rm{Jup}}$ masses), 26 have known massive companions exterior to 200 days (NASA Exoplanet Archive, 2020 August 18). Three of these systems (HIP 14810, \citealt{Wri07}; HIP 57274, \citealt{Fis12}; 55 Cnc, \citealt{But97}) also contain interior companions, making the high-eccentricity migration model unlikely. Since the inclinations of the other 23 companion planets are unknown, we cannot directly simulate the eccentricity evolution of the warm Jupiters, which then prevents us from assessing the viability of perturber-coupled high-eccentricity migration in these systems. However, if this mechanism is common, a uniform RV study targeting warm Jupiter systems over a long time baseline should reveal more massive companions in these systems. Note that all 26 of the warm Jupiter systems with massive exterior companions pass the minimum perturber strength requirement (equation \ref{eq:dong}) outlined in Section \ref{perturbers}.

\citet{Bry16} did conduct a uniform RV survey of 123 exoplanet systems to search for long-period companions through which they found massive companions within 20 au in 8 of their systems as well as long-term RV trends indicative of the presence of long-period companions in an additional 20 systems. However, most of these systems were not warm Jupiter systems. A full analysis of the companions to only the warm Jupiter systems from this study is beyond the scope of this paper, but their overall results still provide useful context. 

There are 7 transiting warm Jupiters with massive companions exterior to 200 days discovered via TTVs and RVs. Of these, 2 systems (Kepler-56, \citealt{Bor11}; Kepler-88, \citealt{Nes13}) contain interior companions making them unsuitable for high-eccentricity migration. Two others, however, have their perturber inclinations constrained well enough to directly simulate the eccentricity evolution of the warm Jupiters (Kepler-419, \citealt{Daw12}; Kepler-693, \citealt{Mas17}). Of these, Kepler-419b is unlikely to have engaged in perturber-coupled high-eccentricity migration unless another undetected object is also perturbing it \citep{Daw14a,Jac19}, while this migration mechanism is plausible for Kepler-693b.

As of writing, no massive companions exterior to 200 days have been found around \textit{TESS} warm Jupiters, but RV follow up efforts will have more targets available than for \textit{Kepler} due to the brightness of the host stars. In fact, RV followup of the newly discovered eccentric warm Jupiter, TOI-677 b \citep{Jor20} shows evidence of a linear trend that may suggest the presence of a long-period companion. If perturber-coupled migration is common, we should expect \textit{TESS} RV followup to discover many such companions (see Section \ref{rv}).

We note that many of the most eccentric warm Jupiters do have known companions (see Figure \ref{fig:ecc}). This trend is discussed by \citet{Soc12} and \citet{Bry16} and may suggest that many of the low-eccentricity warm Jupiters are delivered through other methods.

\section{Discussion and Conclusions}\label{conc}

In this work, we synthesized three populations of warm Jupiter systems: a \textit{Kepler}-like population, a \textit{TESS}-like population, and an RV-discovered population (Sections \ref{kepler_warm}, \ref{tess_warm}, and \ref{rv_warm}). In each warm Jupiter system, we added a stable companion planet with enough perturbing strength to excite large eccentricities in the warm Jupiter in certain orbital configurations (Section \ref{perturbers}). These large eccentricities could have resulted in high-eccentricity tidal migration that delivered the warm Jupiters to their current orbital separations. We then calculated a set of observables (consisting of transit timing variations, transit duration variations, radial velocities, and astrometric signals) that would be induced by the perturbing companion in these systems and compared them with current and near-future observational limits to determine their detectability (Section \ref{results}). We then assessed the sensitivity of these results to our choices for the perturber parameter distributions and reran our calculations with a population of warm Jupiters and perturbers from simulations (Sections \ref{pop_comp} and \ref{petro}). Lastly, we discussed the current list of known massive companions to warm Jupiters in the context of the predictions from our calculations (Section \ref{companions}). We find that many perturbing companions should be detectable in \textit{TESS} and RV warm Jupiter systems in the near future if perturber-coupled high-eccentricity migration is common.

Our specific findings are as follows. If we assume that all warm Jupiters have a perturbing companion that can induce significant eccentricity oscillations leading to high-eccentricity migration, $\sim20\%$ of these companions in \textit{Kepler} warm Jupiters systems should be detectable via TTVs and TDVs in the \textit{Kepler} light curves (Sections \ref{ttv} and \ref{tdv}). With the same assumptions, companions to \textit{TESS} warm Jupiters should be detectable with just 3 months of high-precision RV data in $\sim80\%$ of systems around quiet stars and by their \textit{Gaia} astrometric signal in $\sim15\%$ of systems within 200 pc (Sections \ref{rv} and \ref{astro}). With the same assumptions, companions to RV-discovered warm Jupiters should be detectable with 3 months of high-precision RV data in $\sim30\%$ of systems around quiet stars and by their by their \textit{Gaia} astrometric signal in $\sim30\%$ of systems within 50 pc (Section \ref{rv} and \ref{astro}). Between RV followup and \textit{Gaia} astrometry, a large fraction of the allowed parameter space for strongly perturbing companions in \textit{TESS} and RV-discovered warm Jupiter systems is detectable (Section \ref{detect}). These results are largely robust to the particular parameter distribution from which the perturbers are synthesized (Section \ref{pop_comp}) and the detectability is improved for companions to RV-discovered planets if we consider only short-period warm Jupiters (Section \ref{wj_param}). Lastly, if we run the same calculations on warm Jupiters and perturbers generated by simulations (see \citealt{Pet16}), we find that most of these perturbers (1 $m_{\rm{Jup}}$ planets at 5 au) are not detectable with 3 months of RV data, Gaia astrometry at 200 pc, or \textit{Kepler}-like light curves, but nearly all are detectable with 3 years of RV measurements or \textit{Gaia} astrometry at 50 pc (Section \ref{petro}).

Our results suggest that if the perturber-coupled high-eccentricity migration mechanism is a common delivery pathway for warm Jupiters, many massive companions to these planets should soon be detectable with RV followup of \textit{TESS} and RV-discovered warm Jupiters as well as \textit{Gaia} astrometry. \citet{Bry16} surveyed for companions to RV-discovered short-period planets, but did not distinguish between warm Jupiters and other planets. A reanalysis of their sample considering only the warm Jupiter systems may begin to reveal trends for warm Jupiter companions, but the sample size would be small. The \textit{TESS} mission, however, will soon provide a much larger sample of warm Jupiters suitable for RV followup. A targeted RV search for companions in these systems, even over just a 3 month baseline, should provide a large enough sample to assess the consistency of the observations with our results. \textit{Gaia} astrometry will also add a significant number of systems to that sample with nearby stellar hosts. If the perturber-coupled high-eccentricity migration model is common in the universe, a search for massive companions in these systems should be fruitful. If not, such a search would set a hard upper-limit on the number of warm Jupiter systems that can be explained by this mechanism. \citet{Pet16} argue that, based on their simulations, only $\sim20\%$ of warm Jupiters can be explained by this process and \citet{Daw15} found a dearth of the super-eccentric warm Jupiters predicted by high-eccentricity tidal migration. If only the subset of warm Jupiters that resemble those produced by \citet{Pet16}'s simulations arrived through perturber-coupled high-eccentricity migration, a longer RV time baseline would be needed to find their perturbers.

Alternatives to high-eccentricity tidal migration are that warm and hot Jupiters may have arrived by disk migration \citep{Gol80,War97,Bar14} or formed in situ (e.g., \citealt{Bat16,Bol16}. Recent results by \citet{Bec17} show that all 6 of the transiting hot Jupiters with known exterior companions in their sample must have low mutual inclinations, thus disfavoring Kozai-Lidov migration for hot Jupiters. \citet{Lai18}, however, argue that the spin-orbit coupling parameter is very sensitive to semi-major axis and that the \citet{Bec17} result can not be generalized for all hot Jupiters with companions. \citet{Hua16} disfavor significant migration of any kind for warm Jupiters because they tend to have more sub-jovian companions than their hot Jupiter counterparts. This is problematic because the large observed eccentricities of some warm Jupiters seem to be incompatible with in situ formation since eccentricity excitation from planet-planet scattering is limited by $\nu_{\rm{escape}}/\nu_{\rm{keplarian}}$ \citep{Gol04,Ida13,Per14}. Recent studies have shown, however, that the challenge of reaching large eccentricities on short period orbits may be easier to overcome in systems with 3 or more giant planets \citep{Fre19,And20}. These alternative explanations may be necessary to explain the presence of intermediate eccentricity warm Jupiters if  RV and astrometric followup to \textit{TESS} warm Jupiters does not yield significant numbers of massive companions, as we have shown that such companions are detectable if they exist and are producing intermediate eccentricity warm Jupiters by perturber-coupled high-eccentricity migration.

\acknowledgments
We thank the anonymous referee for their helpful comments on this paper. We gratefully acknowledge support from grant NNX16AB50G awarded by the NASA Exoplanets Research Program and the Alfred P. Sloan Foundation's Sloan Research Fellowship. Computations for this research were performed on the Pennsylvania State University’s Institute for Computational and Data Sciences’ Roar supercomputer. The Center for Exoplanets and Habitable Worlds is supported by the Pennsylvania State University, the Eberly College of Science, and the Pennsylvania Space Grant Consortium. AS received funding from the European Research Council under the European Community’s H2020 2014-2020 ERC Grant Agreement No. 669416 $``Lucky Star"$.  CP acknowledges support from the Bart J. Bok fellowship at Steward Observatory and from ANID – Millennium Science Initiative – ICN12\_009. This research has made use of the NASA Exoplanet Archive, which is operated by the California Institute of Technology, under contract with the National Aeronautics and Space Administration under the Exoplanet Exploration Program.

\software{\textsc{rvlin} \citep{Wri09}, \textsc{mpfit} \citep{Mar09}}

\bibliography{astrobib}

\end{document}